\documentclass{article}

\usepackage{microtype}
\usepackage{graphicx}
\usepackage{subfigure}
\usepackage{booktabs}
\usepackage{hyperref}
\usepackage{amsmath}
\usepackage{amssymb}
\usepackage{mathtools}
\usepackage{amsthm}
\usepackage[capitalize,noabbrev]{cleveref}
\usepackage[textsize=tiny]{todonotes}

\usepackage[accepted]{icml2025}

\usepackage{dsfont}
\usepackage{bm}

\theoremstyle{plain}

\theoremstyle{definition}

\theoremstyle{remark}

\icmltitlerunning{
Challenges and Guidelines in Deep Generative Protein Design: Four Case Studies
}

\begin{document}

\twocolumn[
\icmltitle{
Challenges and Guidelines in Deep Generative Protein Design: Four Case Studies
}

%%%%%%%%%%%%%%%%%%%%%%%%%%%%%%%%
% AUTHORS
%%%%%%%%%%%%%%%%%%%%%%%%%%%%%%%%

% \icmlsetsymbol{equal}{*}

\begin{icmlauthorlist}
\icmlauthor{Tianyuan Zheng}{cam-mathphy}
\icmlauthor{Alessandro Rondina}{unibs-mtmed}
\icmlauthor{Gos Micklem}{cam-genetic}
\icmlauthor{Pietro Liò}{cam-compsci}
\end{icmlauthorlist}

\icmlaffiliation{cam-mathphy}
{Department of Applied Mathematics and Theoretical Physics, University of Cambridge, Cambridge, UK}
\icmlaffiliation{cam-compsci}
{Department of Computer Science and Technology, University of Cambridge, Cambridge, UK}
\icmlaffiliation{unibs-mtmed}
{Department of Molecular and Translational Medicine, University of Brescia, Brescia, Italy}
\icmlaffiliation{cam-genetic}
{Department of Genetics, University of Cambridge, Cambridge, UK}

\icmlcorrespondingauthor{Tianyuan Zheng}{tz365@cam.ac.uk}

%%%%%%%%%%%%%%%%%%%%%%%%%%%%%%%%
% KEY WORDS
%%%%%%%%%%%%%%%%%%%%%%%%%%%%%%%%

\icmlkeywords{Generative Protein Design, Score Matching, Flow Matching, Structural Phylogenetics, Molecular Dynamics, Ligand Docking}

\vskip 0.3in
]

% \printAffiliationsAndNotice{} % leave blank if no need to mention equal contribution
\printAffiliationsAndNotice{\icmlEqualContribution} % otherwise use the standard text.

%%%%%%%%%%%%%%%%%%%%%%%%%%%%%%%%
% ABSTRACT
%%%%%%%%%%%%%%%%%%%%%%%%%%%%%%%%

\begin{abstract}
Deep generative models show promise for \textit{de novo} protein design, yet reliably producing designs that are geometrically plausible, evolutionarily consistent, functionally relevant, and dynamically stable remains challenging.
We present a deep generative modeling pipeline for early \textit{de novo} design of monomeric proteins, based on Score Matching and Flow Matching.
We apply this pipeline to four diverse protein families with an adaptable evaluation protocol.
Generated structures display realistic, clash‐free conformations enriched with family-specific features, while the designed sequences preserve essential functional residues while retaining variability. Molecular dynamics and binding simulations show dynamic stability, with wild-type-like binding pockets that interact favorably with family-specific ligands. 
These results provide practical guidelines for integrating generative models into protein design workflows.
\end{abstract}

%%%%%%%%%%%%%%%%%%%%%%%%%%%%%%%%
% MAIN: Introduction
%%%%%%%%%%%%%%%%%%%%%%%%%%%%%%%%

\section{Introduction}

Protein functions and specificities are dictated by their complex structures. Over the past 60 years, we have progressed from viewing protein design as unattainable to achieving complete artificial design and synthesis \cite{Korendovych2020, Huang2016}, with expanding applications across industries \cite{Arunachalam2021, Kingwell2024, Barclay2023, VictorinodaSilvaAmatto2021, Ali2020}. However, achieving atomic-level precision remains challenging due to the nonlinear complexity of folding and the sensitivity of the function to slight changes, still demanding significant resources.

With the rapid growth of protein structure databases \cite{https://doi.org/10.48550/arxiv.2406.13864, Berman2000}, various \textit{in silico} \textit{de novo} protein design methods, particularly deep generative models \cite{Watson2023, Ingraham2023, https://doi.org/10.48550/arxiv.2209.15611}, have emerged. However, many methods generate structures that appear novel, diverse, and designable, yet few assess their functionality or evolutionary relevance, leaving the observations potentially as mere \cite{Ji2023}. Unclear biological functionality, unknown stability, and the lack of validation connecting these structures to known functions limit the broader application and advancement of these methods.

We tackle these challenges using a deep generative pipeline, based on diffusion-based Score Matching (SM) and Flow Matching (FM), for early \textit{de novo} design of monomeric proteins. Guided by an adaptable evaluation protocol, we apply it to four protein families chosen for their diverse structural folds, functional roles, and rich annotations (\cref{appendix:C}).
Generated structures are realistic and clash-free, and structural phylogenetic analyses show that they capture family-specific features consistent with common ancestry and function.
Designed sequences conserve key functional residues while allowing variability elsewhere, yielding low-similarity variants with similar functions.
Molecular Dynamics (MD) simulations demonstrate dynamic stability of the designs, while forming wild-type (WT)-like binding pockets that favorably interact with family-specific ligands in docking studies. 
Together, these results offer practical guidelines for integrating generative models into protein design workflows.

\section{Pipeline and Evaluation Protocol}

\subsection{Protein Backbone Generation}

In \cref{appendix:B}, we review the key concepts behind the approaches of \citet{FrameDiff} and \citet{FoldFlow} for backbone generation, using SM and FM on $\mathrm{SE}(3)$.

\subsection{Geometric Plausibility}

Due to steric hindrance and spatial repulsion, not all backbone dihedral angles in proteins are physically feasible or energetically favorable. We analyzed the $\phi$ and $\psi$ distribution in the generated structures using Ramachandran plots \cite{Ramachandran1963} (\cref{fig:A.angles}).

\subsection{Conserved Residue Consistency}
\label{sec:conserved_residue_consistency}

In protein families or across species, certain residues are highly conserved, typically to ensure function but also structural stability and to support proper folding.

\paragraph{Determining the optimal amino acid sequence.}

Following \citet{FrameDiff}, we used ProteinMPNN \cite{Dauparas2022} to predict ten sequences for each generated backbone. We then modeled these sequences with EMSFold \cite{rives2019biological} and compared them to the generated backbone using the $\text{TM}_\text{score}$ \cite{Zhang2005}. The sequence with the highest $\text{TM}_\text{score}$ was chosen as the optimal match.

\paragraph{Identifying conserved residues.}

Experimentally derived sequences were aligned with optimal sequences of generated backbones using Clustal Omega \cite{Sievers2011}. Given these alignments and generated structures, ConSurf \cite{Yariv2023} reconstructed phylogenetic trees and applied Rate4Site \cite{Pupko2002} to estimate per-residue evolutionary rates via an empirical Bayesian method \cite{Ashkenazy2016, Mayrose2004}. 

\subsection{Structural Phylogenetics}

Certain applications require designing structures with low sequence similarity while retaining similar functions. However, when sequence identity falls below 30\%, homology detection and evolutionary inference become challenging \cite{PuenteLelievre2023}. Structural comparisons, which are more conserved, offer a more effective alternative \cite{Illergrd2009, Flores1993, Moi2023}.

\paragraph{Using $Q_\text{score}$ in structural phylogenetics.}

\citet{Malik2020} proposed the $Q_\text{score}$ \cite{Krissinel2004} for structural phylogenetics, which accommodates indels and combines alignment quality with length. It compares the positions of all $\mathrm{C}_\alpha$ atoms across $N_\text{align}$ comparable residues in pairwise comparisons. We construct structural phylogenetic trees using $1 - Q_\text{score}$ as a distance measure, where higher values indicate greater structural similarity.

\paragraph{Using 3Di alphabet in structural phylogenetics.}

\citet{vanKempen2023} developed Foldseek, which encodes protein tertiary interactions using a 20-state 3D interaction (3Di) alphabet to simplify structural alignments. This approach reduces false positives and increases information density by effectively encoding conserved core regions. Leveraging Foldseek’s divergence metrics, \citet{Moi2023} constructed structural phylogenetic trees based on rigid body alignment, local alignment without superposition, and sequence alignment with structural alphabets, showing that these trees outperform traditional sequence-based trees across varied evolutionary timescales \cite{Moi2023}.

\subsection{Molecular Dynamics}

Structures that appear reasonable may, in fact, be unstable due to molecular dynamics, water interactions, and entropy, which are not accounted for during generation. To assess the dynamic stability of these generated structures, we conducted MD simulations under physiological conditions and analyzed their time-dependent behavior.

\paragraph{Homology modeling of side-chains.}
\label{sec:md_homo_side}

We used homology modeling to add side-chains to the protein backbone. After determining the optimal sequence (\cref{sec:conserved_residue_consistency}), we selected template proteins with at least 45\% sequence identity and a $\text{TM}_\text{score}$ of 0.75 or higher using FoldSeek \cite{vanKempen2023}. Sequences were aligned with Clustal Omega, and MODELLER \cite{ali1993} generated possible side-chain conformations using statistical potentials and rotamer libraries, with backbone fixed. We chose the final side-chain arrangement based on lowest energy and minimal steric clashes, verifying model quality with PROCHECK \cite{Laskowski1993, Laskowski1996} and WHAT\_CHECK \cite{Hooft1996}, discarding low-quality models. MODELLER's modeling leverages (a) the input alignment to position side-chains, (b) template structures to set spatial restraints that mimic contacts, (c) knowledge-based rotamer distributions to favor typical angles, and (d) an optimization process.

\paragraph{Simulation.}

Hydrogen atoms were added using Reduce2 from the computational crystallography toolbox \cite{GrosseKunstleve2002}, and MD simulations were performed with GROMACS \cite{Abraham2015}. Proteins were placed in an octahedral box with at least 15 $\mathring{\mathrm{A}}$ from the edges, and intermolecular interactions were modeled using the CHARMM36 force field (July 2022) \cite{Vanommeslaeghe2009, Vanommeslaeghe2012, Yu2012}. After vacuum energy minimization, the system was solvated, neutralized to 150 mM Na$^+$ and Cl$^-$, and minimized again. Then it was heated to 310 K under NVT\footnote{Constant number of particles, volume, and temperature.}, equilibrated at 1 atm under NPT\footnote{constant number of particles, pressure, and temperature}, and subjected to a 10 ns production run. Details are provided in \cref{appendix:I}.

\subsection{Protein-ligand Docking}

AutoDock Vina \cite{Trott2009, Eberhardt2021} was used to predict optimal binding modes between generated structures and their family-specific ligands. We performed blind docking, scanning the entire protein surface for potential ligand binding sites without prior knowledge of the pockets. The grid box covered the entire protein, and identical configurations were applied to both generated and experimentally derived structures to evaluate whether deviations in the generated samples fell within an acceptable range, thus assessing their functional viability. Details on simulation settings are provided in \cref{appendix:I3}.

\section{Case Studies}

\subsection{Data}

We assembled a dataset of monomeric proteins covering four families ($\beta$-lactamases, cytochrome \textit{c}, GFP, and Ras), with varied fold types and functions, incorporating both natural and engineered mutations while retaining conserved core functional regions. Details are provided in \cref{appendix:C}.

\subsection{Backbone Generation}

We trained the SM and FM (with Optimal Transport) models on these protein families, using pretrained weights from \citet{FrameDiff} and \citet{FoldFlow}, with each model having $\sim 17$ million parameters. Each model generated 50 backbone structures, with target sequence lengths sampled from the distribution observed in the training data (\cref{fig:A.aadist}).

\subsection{Backbone Dihedral Angles}

In \cref{fig:A.angles}, most data points fall within the allowed and favored regions, with few in the disallowed areas, and no significant geometric clashes or unreasonable conformations observed. The dihedral angle distributions of the generated structures align well with those of the experimentally derived proteins used for training. For instance, cytochrome \textit{c} structures have sparse points in the region $-180^\circ < \psi < -90^\circ$ and $45^\circ < \phi < 180^\circ$, while GFP structures form four clusters there. SM samples are closer to the training data and concentrate in allowed regions and show less diversity than FM samples.

\subsection{Conserved Residue Consistency}

The optimal sequences preserved conserved residues that were largely consistent with the experimentally derived data (\cref{fig:A.resevo}).
In \cref{fig:A.resevs}, similar to the experimentally derived proteins, the generated structures have increased residue conservation around binding pockets (refer to \cref{fig:A.proteins} and \cref{fig:A.mddocking}). For instance, in cytochrome \textit{c} (1HRC) and generated SM-1 and FM-0, conserved residues cluster near the central heme C binding site. A similar pattern appears in 4OBE and the generated KRas proteins.

\subsection{Structural Phylogenetic Tree}

\paragraph{Summary tree.}

Pairwise structural distances between generated and experimental structures were computed using two methods, forming a matrix for phylogenetic tree construction. After normalizing branch lengths, a summary tree was generated with SumTrees \cite{dendropy5} (\cref{fig:A.phylo.summary}). The topology shows that both the generated and the experimentally derived proteins cluster strongly by family, without intermixing among these groups.

\paragraph{Structure-informed tree outperforms sequence-only tree.} 

Sequence-based phylogenetic trees were constructed by aligning sequences with Clustal Omega and inferring trees with FastTree \cite{Price2009}. Structure trees better preserve natural taxonomic groupings than sequence trees (\cref{fig:A.phylo}AB), especially at moderate to low sequence identity, and they achieve higher Taxonomic Congruence Scores \cite{Tan2015} (\cref{appendix:G}, \cref{fig:A.phylo}D), indicating closer alignment with the known taxonomy (\cref{appendix:G}).

\subsection{Dynamic Stability}

Dynamic stability was analyzed as:
(1) 
Backbone RMSD over time, with stable proteins typically below $2 \mathring{\text{A}}$, or up to $3 \mathring{\text{A}}$ for larger, flexible proteins \cite{Burton2012, Liu2017, Wong2024}. In \cref{fig:A.mdmetrics}A, experimentally derived structures remain within $2 \mathring{\text{A}}$, while generated structures average $\sim 2.5 \mathring{\text{A}}$, rarely exceeding $3 \mathring{\text{A}}$. SM samples show RMSD $\sim 0.3 \mathring{\text{A}}$ higher than FM samples.
(2)
Radius of gyration ($R_g$) quantifies the spatial distribution of a molecule's atoms relative to its center of mass. Generated structures are expected to be compact, with $R_g$ values close to experimentally derived structures and fluctuations around $1 \mathring{\text{A}}$ (\cref{fig:A.mdmetrics}B).
(3)
The DSSP algorithm assigns secondary structure to each residue based on hydrogen bonds and geometry. In \cref{fig:A.mdmetrics}D, the stability in secondary structure elements over time suggests structural stability with no major conformational changes.
(4)
Lower potential energy are generally more stable
, with contributions from bond, angle, dihedral, van der Waals, and electrostatic energies. In \cref{fig:A.mdmetrics}C, most generated samples have lower potential energy than the experimentally derived data.

\subsection{Protein-ligand Docking}

Docking on experimentally derived structures closely matches known binding modes, with ligand RMSDs $\sim 1.5 \mathring{\text{A}}$ and low binding energies (\cref{fig:A.mddocking}). Successful docking typically shows binding free energies ($\Delta G$) between -7 and -10 kcal/mol, with lower values indicate stronger, more stable interactions \cite{P2021, Nguyen2019}. Generated structures also have pockets similar to those of experimentally derived proteins (\cref{fig:A.mddocking}). In most simulations, ligands bind within $4 \mathring{\text{A}}$ of experimentally derived positions with $\Delta G$ below -6 kcal/mol. SM samples generally show lower $\Delta G$ and RMSDs than FM samples.

\paragraph{$\beta$-lactamases binding to penicillin.}

Mutations at Glu$^{166}$ and Asn$^{170}$ in class A $\beta$-lactamases (\cref{fig:A.proteins}A) can form a stable acyl-enzyme intermediate, disrupting deacylation \cite{Chen2001}. Only Asn and Gly at 170 preserve WT-like function \cite{Brown2009}, and this conservation is retained in generated samples like FM-4 (\cref{fig:A.mddocking}A), suggesting they may retain the ability to inactivate penicillin antibiotics. Notably, Asn/Gly at position 170 occurred in 23/50 FM samples and 18/50 SM samples, and when extended to positions 169 to 171, in 30/50 FM samples and 22/50 SM samples, rates higher than expected by chance.

\paragraph{Cytrochrome \textit{c} binding to heme c.}

The generated cytochrome c-like structures retain conserved residues found in the WT (\cref{fig:A.proteins}B; 1HRC). SM-1 (\cref{fig:A.mddocking}B) includes phosphorylatable residues Tyr$^{58}$, Thr$^{59}$, and Tyr$^{107}$, as well as lysine residues Lys$^{82}$, Lys$^{83}$, and Lys$^{96}$ involved in phospholipid binding. Asn$^{80}$ and these lysines form an ATP-binding pocket-like structure. FM-0 shows heme iron coordinated by two cysteines, which may form stronger covalent bonds, potentially affecting electron transfer efficiency.

\paragraph{KRas binding to GNP/GDP.}

GNP, a non-hydrolyzable GTP analog, is commonly used to simulate the GTP-bound active state of Ras proteins \cite{Pantsar2020}. In WT KRas bound to GNP (\cref{fig:A.mddocking}D; 5UFE), the Switch II region residues fill the pocket. Removing of the $\gamma$-phosphate relaxes the conformation to the GDP-bound state, opening the switch II region (\cref{fig:A.mddocking}C; 4OBE) \cite{Kauke2017}.

After blind docking the generated KRas-like structures with GNP/GDP, we analyzed the complexes using protein–ligand MD simulations. Allowing flexibility in both the backbone and side-chains enabled us to capture the dynamic conformational changes, especially the WT-like transitions in the switch regions, between active/inactive states. 
FM-generated structures show greater flexibility. In FM-13 (\cref{fig:A.mddocking}D), GNP binding causes switch II residues to fill the pocket, similar to 5UFE. In the GDP-bound state, FM-13 adopts an open conformation, with the channel between the switches and P-loop opening, as seen in 4OBE (\cref{fig:A.mddocking}C). In contrast, SM-27 are more rigid, with fewer conformational changes between GDP- and GTP-bound states.

\section{Discussion}

SM and FM can generate a range of monomeric protein structures and can have applications beyond protein design \cite{FrameDiff, FoldFlow}. However, the complexity of these tasks is often underestimated, and function verification for novel samples remains costly. Targeting specific tasks or integrating generative methods into well-established empirical knowledge may yield better results. Key factors like conformational dynamics, water interactions, and entropy have not been fully considered in generation \cite{Du2024}. Incorporating protein sequences, side-chain details, and functional annotations as context could improve model performance. Although some progress has been made \cite{https://doi.org/10.48550/arxiv.2308.07416, https://doi.org/10.48550/arxiv.2301.10814, https://doi.org/10.48550/arxiv.2302.11419, Zhou2023}, research gaps remain.

The generated samples for the most part capture the observed evolutionary diversity. However, there are regions of the observed diversity that are not so well covered (\cref{fig:A.phylo.summary}) (such as GFP and class A $\beta$-lactamases), raising concerns about potential overfitting and limited generalizability. Common metrics are inadequate for quantifying overfitting in generative models \cite{https://doi.org/10.48550/arxiv.1703.00573}, especially given the variability in sample sizes and quality across families.

\section{Guidelines}

Here, we outline a set of guidelines and best practices for designing and validating proteins using deep generative methods: 
(1) Implement physical constraints during the generation process to produce realistic structures without severe steric clashes.
(2) Generate multiple candidate sequences for each design and prioritize those most likely to fold into the target structure. 
(3) Integrate sequence design and backbone validation early to ensure compatibility between the sequence and the backbone model.
(4) Identify and preserve conserved residues in the designs to maintain critical functional motifs and overall structural integrity.
(5) Construct all-atom models of the designs and refine them to relieve any strain or clashes introduced during initial modeling.
(6) Compare the designs to known protein family structures or fold templates to ensure consistency with known protein folds.
(7) Perform molecular dynamics simulations to verify that the proposed structures remain stable under physiologically relevant conditions.
(8) Perform docking analyses to evaluate the viability of proposed ligand-binding sites.
(9) Experimental validation is always essential to confirm the intended function and structural integrity of each candidate.
(10) See \cref{appendix:J} for circumstances under which implementation of these guidelines can be challenging.

\section{Conclusion}

This study shows the potential of methods based on deep generative models for designing proteins with high structural fidelity and functional plausibility. SM better captures conserved regions, producing more rigid structures, while FM offers greater flexibility (\cref{fig:A.resevo}). Although the optimized sequences for the generated structures exhibit some differences from natural family members, they preserve many functionally essential conserved residues (\cref{fig:A.resevs}). In structural phylogenetic trees (\cref{fig:A.phylo.summary}), these designs cluster with their respective families, suggesting that their geometries could capture some family‐specific signatures indicative of common ancestry or function. MD and docking simulations are consistent with the designs remain properly folded and stable under physiological conditions over time (\cref{fig:A.mdmetrics}), and can accommodate their family-specific ligands in binding pockets that closely correspond to known active sites (\cref{fig:A.mddocking}). Some designs undergo conformational changes upon binding that mirror the allosteric behavior observed in WT (\cref{fig:A.mddocking}CD).

\section*{Software and Data}

All input data and software used in this study are available from public sources or provided under academic licenses. The source code, scripts, generated samples, and curated datasets can be accessed at \url{https://github.com/ECburx/PROTEVAL}. The PDB structure files were downloaded on June 19, 2024, from \url{https://www.wwpdb.org/ftp/pdb-ftp-sites}. 

\section*{Acknowledgements}

The authors would like to thank Charlie Harris, Simon Mathis and Stephen Eglen for their fruitful discussions and valuable feedback on this work.

%%%%%%%%%%%%%%%%%%%%%%%%%%%%%%%%
% BIBLIOGRAPHY
%%%%%%%%%%%%%%%%%%%%%%%%%%%%%%%%

% In the unusual situation where you want a paper to appear in the references without citing it in the main text, use \nocite
% \nocite{langley00}

\bibliography{bib}
\bibliographystyle{icml2025}

%%%%%%%%%%%%%%%%%%%%%%%%%%%%%%%%
% APPENDIX
%%%%%%%%%%%%%%%%%%%%%%%%%%%%%%%%

\newpage
\appendix
\onecolumn
\section{Deep Generative Protein Design Workflow}
\label{appendix:A}

\begin{figure*}[!ht]
\begin{center}
\centerline{\includegraphics[width=\textwidth]{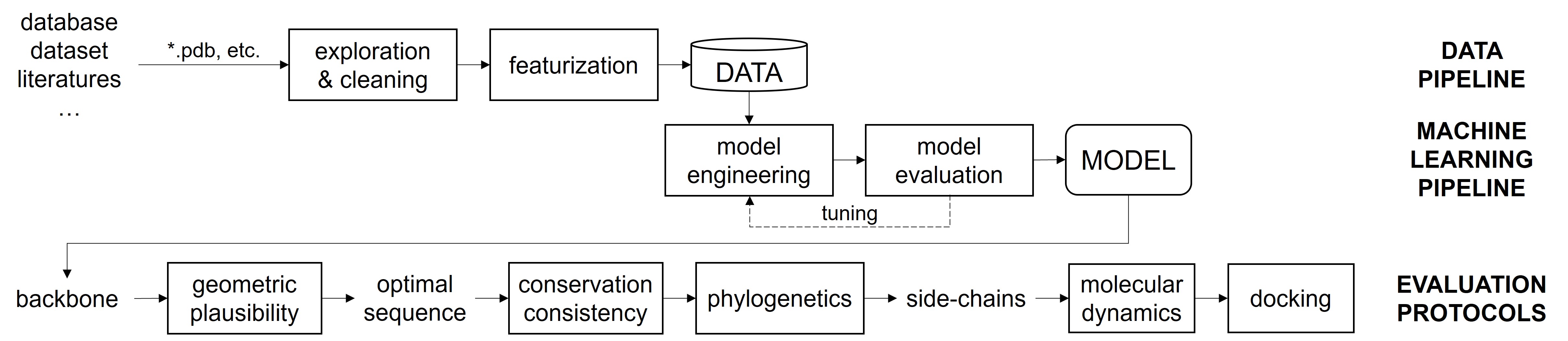}}
\vskip -0.1in
\caption{
A schematic overview of the deep generative protein design pipeline and its evaluation protocols.
}
\label{fig:A.pipeline}
\end{center}
\vskip -0.2in
\end{figure*}

\newpage
\section{Protein Backbone Generation}
\label{appendix:B}

\subsection{$\mathrm{SE}(3)$ Decomposition into $\mathrm{SO}(3)$ and $\mathbb{R}^3$}
\label{appendix:B1}

The Special Euclidean group $\mathrm{SE}(3)$ describes the rotations and translations in 3D space. An element of $\mathrm{SE}(3)$ can be represented by a $4 \times 4$ matrix:
\begin{align}
\bm{\mathsf{T}} = \begin{pmatrix}
\bm{\mathsf{R}}         & \bm{\mathsf{x}} \\
\mathbf{0}_{1 \times 3} & 1
\end{pmatrix}
\end{align}
where $\bm{\mathsf{R}}$ is a $3\times 3$ rotation matrix belonging to the Special Orthogonal group $\mathrm{SO}(3)$, and $\bm{\mathsf{x}} = [\mathsf{x}_x \ \mathsf{x}_y \ \mathsf{x}_z] \in \mathbb{R}^3$ is the translational component. Since $\mathrm{SE}(3)$ can be viewed as the semidirect product of $\mathrm{SO}(3)$ and $\mathbb{R}^3$, denoted as $\mathrm{SE}(3) \cong \mathrm{SO}(3) \ltimes \mathbb{R}^3$, one option is to naturally treat $\mathrm{SO}(3)$ and $\mathbb{R}^3$ as independent for simplicity \cite{FrameDiff}.

\subsection{Protein Backbone Representations}
\label{appendix:B2}

\begin{figure*}[!ht]
\begin{center}
\centerline{\includegraphics[width=.35\textwidth]{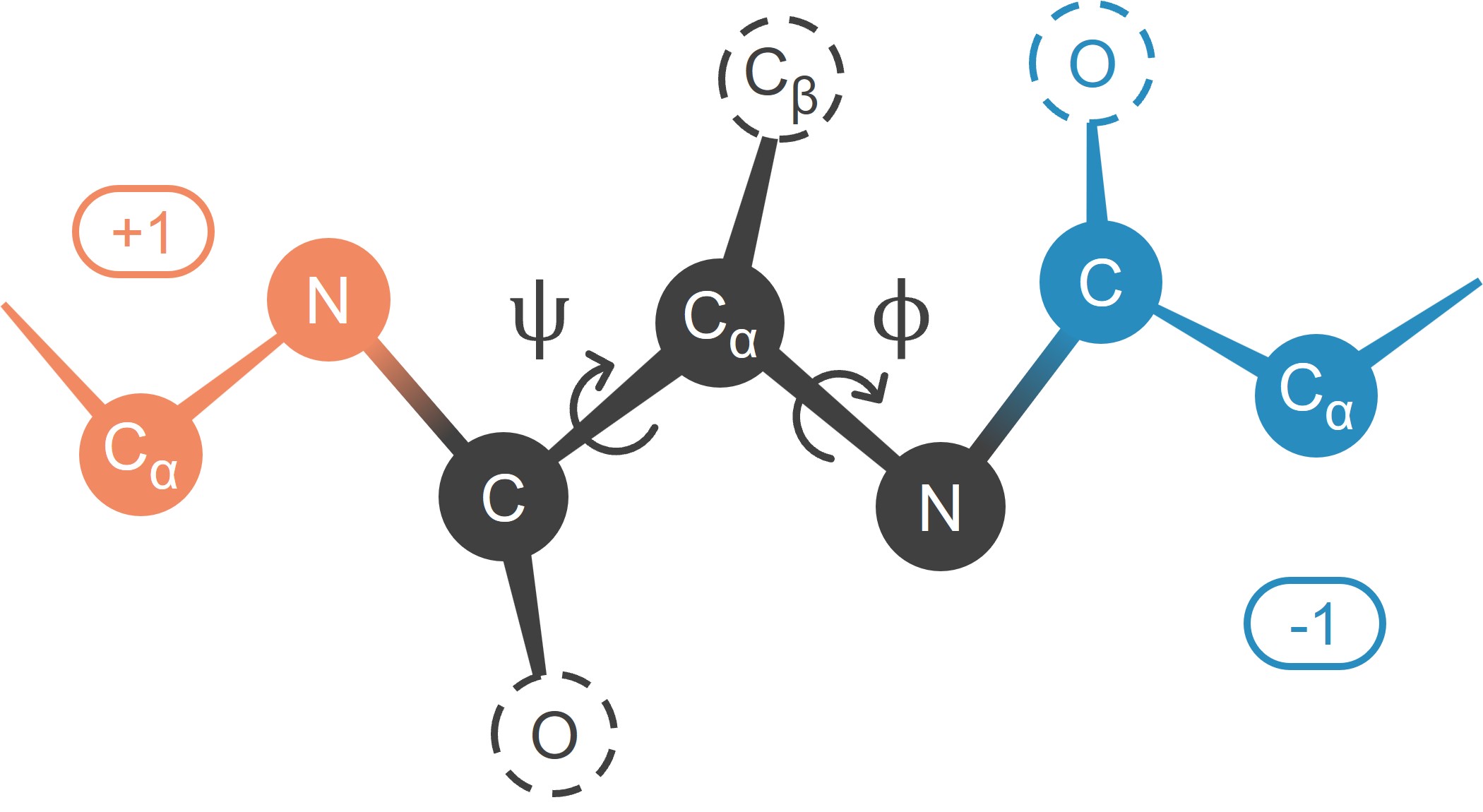}}
\vskip -0.1in
\caption{
Protein backbone with dihedral angles $\psi$ and $\phi$. 
}
\label{fig:A.rep}
\end{center}
\vskip -0.2in
\end{figure*}

Molecules can be intuitively represented as 3D atomic point clouds. However, macromolecules like proteins may contain thousands or tens of thousands of atoms, with variation in the atom types and quantities among different amino acids (for instance, sulfur atoms are present only in a few amino acids like cysteine). Representing proteins as unordered 3D atomic point clouds significantly increases data dimensionality and sparsity, requiring far more training data than is typically available.

Following the work of \citet{FrameDiff} and \citet{FoldFlow}, we adopt the more compact backbone rigid groups from AlphaFold \cite{AlphaFold2} to represent protein backbone structures in 3D space. A backbone rigid group consists of the main chain atoms (N, C$_\alpha$, C, O) within a single residue (\cref{fig:A.rep}), where their geometric relationships (relative positions and orientations) are highly stable. The position and orientation of the group is transformed as a whole, without accounting for individual atomic movements, simplifying the computation and reducing structural errors caused by excessive degrees of freedom.

Assuming experimentally derived ideal chemical bond angles and lengths, models learn how the rigid transformation (or \textit{frame}) $\bm{\mathsf{T}}_i$ of each residue $i \in [1, N]$ acts on idealized coordinates $[\mathrm{N}^{\star}, \mathrm{C}_\alpha^{\star}, \mathrm{C}^{\star}] \in \mathbb{R}^3$ (centered at $\mathrm{C}_\alpha^{\star}$), so that the transformed coordinates match the actual coordinates as closely as possible:
\begin{align}
[\mathrm{N}, \mathrm{C}_\alpha, \mathrm{C}]_i = \bm{\mathsf{T}}_i \cdot [\mathrm{N}^{\star}, \mathrm{C}_\alpha^{\star}, \mathrm{C}^{\star}]
\end{align}
where $\bm{\mathsf{T}}_i \in \mathrm{SE}(3)$ can be decomposed into a rotation matrix $\bm{\mathsf{R}}_i \in \mathrm{SO}(3)$ and a translation vector $\bm{\mathsf{x}}_i \in \mathbb{R}^3$. An additional torsion angle $\psi_i \in \mathrm{SO}(2)$ is introduced between the bond of $\mathrm{C}_\alpha$ and C for a more accurate construction of the backbone oxygen atom O.

\subsection{$\mathrm{SE}(3)$ Score Matching}
\label{appendix:B3}

Let $\mathbf{T}_t = \left[ \bm{\mathsf{T}}_{1,t}, \dots, \bm{\mathsf{T}}_{N,t} \right] \in \mathrm{SE}(3)^N$ denote the manifold of $N$ frames at time $t$, where each frame can independently rotate and translate. Correspondingly, define $\mathbf{R}_t = \left[ \bm{\mathsf{R}}_{1,t}, \dots, \bm{\mathsf{R}}_{N,t} \right]$ and $\mathbf{X}_t = \left[ \bm{\mathsf{x}}_{1,t}, \dots, \bm{\mathsf{x}}_{N,t} \right]$. By treating $\mathrm{SO}(3)$ and $\mathbb{R}^3$ as two independent stochastic processes, a \textit{forward process} gradually perturbs the initial data distribution $p_0$. Following the approach of \citet{FrameDiff}, this process is described by the Stochastic Differential Equation (SDE) for $\mathbf{T}_t \sim p_t$ and any arbitrary time $t \in [0, T]$: 
\begin{align}
\mathrm{d}\mathbf{T}_t
= \left[ 0, -\frac{1}{2} \mathbf{X}_t \right]\mathrm{d}t + \left[ \mathrm{d}\mathbf{B}_t^{\mathrm{SO}(3)}, \mathrm{d}\mathbf{B}_t^{\mathbb{R}^3} \right]
\label{eq:SM-FORWARD}
\end{align}
where $\mathbf{B}_t^{\mathrm{SO}(3)}$ and $\mathbf{B}_t^{\mathbb{R}^{3}}$ are Brownian motions on the $\mathrm{SO}(3)$ and $\mathbb{R}^{3}$, respectively. Invariant density $p_{\mathrm{inv}}^{\mathrm{SE}(3)}(\bm{\mathsf{T}}) \propto \mathcal{U}^{\mathrm{SO}(3)}(\bm{\mathsf{R}}) \ \mathcal{N}(\bm{\mathsf{x}};0,\mathbf{I})$ is chosen for $\bm{\mathsf{T}} = (\bm{\mathsf{R}}, \bm{\mathsf{x}})$. 

Let $(\overleftarrow{\mathbf{T}}_t)_{t\in[0,T]} = (\mathbf{T}_{T-t})_{t\in[0,T]}$, the corresponding \textit{time-reverse process} \cite{https://doi.org/10.48550/arxiv.2202.02763} is given by
\begin{align}
\mathrm{d}\overleftarrow{\mathbf{R}}_t
&= \nabla_{\bm{\mathsf{R}}} \log p_{T-t}(\overleftarrow{\mathbf{T}}_t)\mathrm{d}t + 
\mathrm{d}\mathbf{B}_t^{\mathrm{SO}(3)}
\label{eq:SM-REVERSE-1}
\\
\mathrm{d}\overleftarrow{\mathbf{X}}_t
&= \left\{ 
\frac{ \overleftarrow{\mathbf{X}}_t }{2} + \nabla_{\bm{\mathsf{x}}} \log p_{T-t} (\overleftarrow{\mathbf{T}}_t)
\right\} \mathrm{d}t +
\mathrm{d}\mathbf{B}_t^{\mathbb{R}^{3}}
\label{eq:SM-REVERSE-2}
\end{align}
where $\nabla \log p$ is the gradient of the log-probability density function (also known as the Stein \textit{score}). However, this gradient is typically intractable in practice because the exact form of $p_t(\mathbf{T}_t)$ is unknown at any given time $t$.

Instead, score-based models estimate tractable conditional score $\nabla \log p_{t|0}$ through SM \cite{Vincent2011}, using a neural network $s(\theta, t,\cdot)$ trained by minimizing both\footnote{One adds weights $1/\mathbb{E} [\lVert \nabla_{\mathbf{R}} \log p_{t|0} (\mathbf{R}_t | \mathbf{R}_0) \rVert^2 ]$ to \cref{eq:SM-LOSS-R} and $(1-e^{-t})/e^{-t/2}$ to \cref{eq:SM-LOSS-X} for simplicity.}:
% {\fontsize{9}{10.2}\selectfont
\begin{align}
\mathcal{L}_\text{SM}^\mathbf{R}(\theta) 
&= \mathbb{E}
\left[
\lVert
\nabla_{\mathbf{R}} \log p_{t|0} (\mathbf{R}_t | \mathbf{R}_0) - s(\theta, t, \mathbf{R}_t)
\rVert^2
\right]
\label{eq:SM-LOSS-R}
\\
\mathcal{L}_\text{SM}^\mathbf{X}(\theta) 
&= \mathbb{E}
\left[
\lVert
\nabla_{\mathbf{X}} \log p_{t|0} (\mathbf{X}_t | \mathbf{X}_0) - s(\theta, t, \mathbf{X}_t)
\rVert^2
\right]
\label{eq:SM-LOSS-X}
\end{align}
% }
with $t\sim \mathcal{U}(0,T)$ and
% {\fontsize{8.5}{10.2}\selectfont
\begin{align}
\nabla_{\bm{\mathsf{R}}} \log p_{t|0} (\bm{\mathsf{R}}_t | \bm{\mathsf{R}}_0)
&= 
\frac{\bm{\mathsf{R}}_t}{\omega( \bm{\mathsf{R}}_{0 \to t} )}
\log \{ \bm{\mathsf{R}}_{0 \to t} \} 
\frac
{\partial_\omega f(\omega( \bm{\mathsf{R}}_{0 \to t} ),t)}
{f(\omega( \bm{\mathsf{R}}_{0 \to t} ),t)} 
% \notag
\label{eq:SM-SCORE-R}
\\
\nabla_{\bm{\mathsf{x}}} \log p_{t|0} (\bm{\mathsf{x}}_t | \bm{\mathsf{x}}_0)
&= \frac{e^{-t/2}\bm{\mathsf{x}}_0 - \bm{\mathsf{x}}_t}{1-e^{-t}}
% &= (1-e^{-t})^{-1}(e^{-t/2}\bm{\mathsf{x}}_0 - \bm{\mathsf{x}}_t) 
% \notag
\label{eq:SM-SCORE-X}
\end{align}
% }
where $\omega$ represents the rotation angle, $\bm{\mathsf{R}}_{0 \to t} = \bm{\mathsf{R}}_0^\top \bm{\mathsf{R}}_t$, and 
\begin{align}
f(\omega, t) = \sum_{\ell \in \mathbb{N}} (2\ell + 1) e^{-\ell(\ell + 1)t/2} \frac{\sin((\ell + \frac{1}{2})\omega)}{\sin(\frac{\omega}{2})}
\end{align}
is an auxiliary function for the heat kernel\footnote{The heat kernel on a manifold is the fundamental solution to the heat equation, representing the probability density function of a Brownian particle diffusing from one point to another over time.} of the Brownian motion on $\mathrm{SO}(3)$.

\subsection{$\mathrm{SE}(3)$ Flow Matching}
\label{appendix:B4}

FM is a simulation-free method for training vector fields to follow a prescribed conditional probability path \cite{FlowMatching}. Formally, for $t \in [0, 1]$, let $\mathbf{U} = \{\mathbf{u}_t\}$ be a \textit{flow} which is a set of time-indexed vector fields that describe the paths along which data points move from an initial distribution $p_1$ to a target distribution $p_0$. Each vector field $\mathbf{u}_t(\mathbf{T}_t)$ represents the rate of change of $\mathbf{T}_t$ which is typically the solution to the Ordinary Differential Equation (ODE) $\frac{d}{dt}\mathbf{T}_t = \mathbf{u}_t(\mathbf{T}_t)$. FM approximates $\mathbf{u}_t(\mathbf{T}_t)$ with a network $v(\theta, t,\cdot)$ by minimizing $\mathcal{L}_\text{FM}(\theta) = \mathbb{E} \left\lVert \mathbf{u}_t(\mathbf{T}_t) - v(\theta, t,\mathbf{T}_t) \right\rVert^2$ with $t\sim \mathcal{U}(0,1)$.

Similarly, independent flows can be built on $\mathrm{SO}(3)$ and $\mathbb{R}^3$. Computing $\mathbf{u}_t$, however, is also intractable due to the complex integrals involved in defining the marginal probability path and vector field. By showing $\nabla_\theta \mathcal{L}_\text{FM}(\theta) = \nabla_\theta \mathcal{L}_\text{CFM}(\theta)$, \citet{FlowMatching} suggested the tractable conditional FM objective on $\mathbb{R}^3$:

\begin{align}
\mathcal{L}_\text{CFM}^\mathbf{X}(\theta) 
= 
\mathbb{E} \left\lVert 
\mathbf{u}_t(\mathbf{X}_t|\mathbf{X}_0) 
-
v(\theta, t,\mathbf{X}_t)
\right\rVert^2
\label{eq:FM-R3-1}
\end{align}

with the Gaussian path $p_t(\bm{\mathsf{x}}_t|\bm{\mathsf{x}}_0) = \mathcal{N}(\bm{\mathsf{x}}_t; t \bm{\mathsf{x}}_0, (t\sigma - t + 1)^2 )$ generated by
\begin{align}
\mathbf{u}_t(\bm{\mathsf{x}}_t|\bm{\mathsf{x}}_0) 
=
\frac{\bm{\mathsf{x}}_0-(1-\sigma)\bm{\mathsf{x}}_t}{1-(1-\sigma)t}
\label{eq:FM-R3-2}
\end{align}
where $\sigma > 0$ is a smoothing constant.

For flows on $\mathrm{SO}(3)$, \citet{FoldFlow} set
\begin{align}
\mathcal{L}_\text{CFM}^\mathbf{R}(\theta) 
= 
\mathbb{E} \left\lVert 
\mathbf{u}_t(\mathbf{R}_t|\mathbf{R}_0,\mathbf{R}_1) 
-
\mathbf{v}(\theta, t,\mathbf{R}_t)
\right\rVert^2
\label{eq:FM-SO3-1}
\end{align}
and define the geodesic interpolant between $\bm{\mathsf{R}}_1 \sim p_1$ and $\bm{\mathsf{R}}_0 \sim p_0$ as $\bm{\mathsf{R}}_t = \exp_{\bm{\mathsf{R}}_1}(t \log_{\bm{\mathsf{R}}_1}(\bm{\mathsf{R}}_0))$. Let $\Psi_t$ be a flow that connects $\bm{\mathsf{R}}_1$ to $\bm{\mathsf{R}}_0$, computing $\mathbf{u}_t(\bm{\mathsf{R}}_t|\bm{\mathsf{R}}_0,\bm{\mathsf{R}}_1)$ simplifies to determining $\bm{\mathsf{R}}_t$ along $\frac{\mathrm{d}}{\mathrm{d}t} \Psi_t(\bm{\mathsf{R}}) = \dot{\bm{\mathsf{R}}}_t$ \cite{https://doi.org/10.48550/arxiv.2302.03660} and then taking its time derivative. Thus, we have
\begin{align}
\mathbf{u}_t(\bm{\mathsf{R}}_t|\bm{\mathsf{R}}_0,\bm{\mathsf{R}}_1)
=
\frac{\log_{\bm{\mathsf{R}}_t} (\bm{\mathsf{R}}_0)}{t}
\label{eq:FM-SO3-2}
\end{align}

\paragraph{Optimal transport.}

Optimal Transport (OT) conditions hold when the probability paths between two distributions are defined by a displacement map that linearly interpolates between them \cite{https://doi.org/10.48550/arxiv.2304.14772}. 

\citet{https://doi.org/10.48550/arxiv.2302.00482} views the OT problem as finding a mapping that minimizes the 2-Wasserstein distance between two distributions $p_1$ and $p_0$ on $\mathbb{R}^3$, using the Euclidean distance $\lVert \bm{\mathsf{x}}_0 - \bm{\mathsf{x}}_1 \rVert$ as the displacement cost:
\begin{align}
W(p_0,p_1)^2_2 = \inf_{\pi \in \Pi} \int_{\mathbb{R}^3 \times \mathbb{R}^3} \lVert \bm{\mathsf{x}}_0 - \bm{\mathsf{x}}_1 \rVert^2 \mathrm{d}\pi(\bm{\mathsf{x}}_0, \bm{\mathsf{x}}_1)
\end{align}
where $\Pi$ denotes the set of all joint probability measures on $\mathbb{R}^3 \times \mathbb{R}^3$ with marginals $p_1$ and $p_0$. By setting $p(\bm{\mathsf{x}}_0, \bm{\mathsf{x}}_1) = \pi(\bm{\mathsf{x}}_0, \bm{\mathsf{x}}_1)$ and a Gaussian conditional probability path with mean $\mu_t = t \bm{\mathsf{x}}_0 + (1-t) \bm{\mathsf{x}}_1$, we have
\begin{align}
&\mathcal{L}_\text{OT}^\mathbf{X}(\theta) 
= 
\mathbb{E}_\pi \left\lVert 
\mathbf{u}_t(\mathbf{X}_t|\mathbf{X}_0, \mathbf{X}_1) 
-
v(\theta, t,\mathbf{X}_t)
\right\rVert^2
\\
&\mathbf{u}_t(\bm{\mathsf{x}}_t | \bm{\mathsf{x}}_0, \bm{\mathsf{x}}_1) 
=
\bm{\mathsf{x}}_0 - \bm{\mathsf{x}}_1
\end{align}
with $p_t(\bm{\mathsf{x}}_t) = \int \mathcal{N} (\bm{\mathsf{x}}_t | t \bm{\mathsf{x}}_0 + (1-t)\bm{\mathsf{x}}_1, \sigma^2) \pi(\bm{\mathsf{x}}_0, \bm{\mathsf{x}}_1) \mathrm{d}\bm{\mathsf{x}}_0 \mathrm{d}\bm{\mathsf{x}}_1$.

Inspired by this, \citet{FoldFlow} extended \cref{eq:FM-SO3-1} and \cref{eq:FM-SO3-2} to $\mathrm{SO}(3)$ using Riemannian optimal transport, with $\bar{\pi}$ being the projection of $\pi$ on $\mathrm{SO}(3)$:
\begin{align}
\mathcal{L}_\text{OT}^\mathbf{R}(\theta) 
= 
\mathbb{E}_{\bar{\pi}} \left\lVert 
\frac{\log_{\mathbf{R}_t} (\mathbf{R}_0)}{t}
-
v(\theta, t,\mathbf{R}_t)
\right\rVert^2
\end{align}

\subsection{$\mathrm{SE}(3)$ Invariance}

$\mathrm{SE}(3)$ invariance can be achieved by consistently positioning the model at the origin \cite{FrameDiff, FoldFlow, https://doi.org/10.48550/arxiv.2008.12577}.

In the context of SM, to ensure translation invariance on $\mathbb{R}^3$, one applies a projection matrix $\mathbf{P} \in \mathbb{R}^{3N \times 3N}$ that removes the center of mass $\frac{1}{N} \sum_{i=1}^N \bm{\mathsf{x}}_i$. It results in an invariant measure on $\mathrm{SE}(3)^N$, denoted as $\mathrm{SE}(3)^N_0$. Since the Brownian motion on $\mathrm{SO}(3)$ and the score $\nabla_{\bm{\mathsf{R}}} \log p_{T-t}$ are both rotation-invariant, \cref{eq:SM-REVERSE-1} is $\mathrm{SO}(3)$-invariant. Consequently, \citet{FrameDiff} derive the following $\mathrm{SE}(3)$-invariant forward process
\begin{align}
\mathrm{d}\mathbf{T}_t
= \left[ 0, -\frac{1}{2} \mathbf{P} \mathbf{X}_t \right]\mathrm{d}t + \left[ \mathrm{d}\mathbf{B}_t^{\mathrm{SO}(3)^N}, \mathbf{P}\mathrm{d}\mathbf{B}_t^{\mathbb{R}^{3N}} \right]
\end{align}
and its corresponding time-reverse process
{\fontsize{9.5}{11.4}\selectfont
\begin{align}
\mathrm{d}\overleftarrow{\mathbf{R}}_t
&= \nabla_{\bm{\mathsf{R}}} \log p_{T-t}(\overleftarrow{\mathbf{T}}_t)\mathrm{d}t + 
\mathrm{d}\mathbf{B}_t^{\mathrm{SO}(3)^N}
\\
\mathrm{d}\overleftarrow{\mathbf{X}}_t
&= \mathbf{P} \left\{ 
\frac{ \overleftarrow{\mathbf{X}}_t }{2} + \nabla_{\bm{\mathsf{x}}} \log p_{T-t} (\overleftarrow{\mathbf{T}}_t)
\right\} \mathrm{d}t +
\mathrm{d}\mathbf{P}\mathbf{B}_t^{\mathbb{R}^{3N}}
\end{align}
}The same approach can be applied to FM. After centering and decoupling the flow on $\mathrm{SE}(3)_0^N$, a separate $\mathrm{SE}(3)$-invariant flow can be constructed for each residue in backbone\footnote{As the product group of $N$ copies of $\mathrm{SE}(3)$, $\mathrm{SE}(3)_0^N$ has a geometric structure that allows global geometric operations (such as geodesic distance, exponential maps, and logarithmic maps) to be decomposed into operations on each of the $N$ $\mathrm{SE}(3)$ groups.}, in which each $\mathrm{SE}(3)$-invariant measure is decomposed into a measure that is proportional to the Lebesgue measure on $\mathbb{R}^3$ \cite{Pollard2001} and an $\mathrm{SO}(3)$-invariant measure \cite{FoldFlow}.

\subsection{Additional Losses}

To prevent unrealistic fine-grained features such as steric clashes or chain breaks when learning the torsion angle $\psi$, \citet{FrameDiff} proposed adding two additional loss functions. The first is the mean squared error (MSE) on backbone atom positions:
\begin{align}
\mathcal{L}_\text{bb} =
\frac{1}{4N} \sum_{n=1}^N \sum_{a \in A}
\lVert a_n - \hat{a}_n\rVert^2
\end{align}
where $A = \{\mathrm{N},\, \mathrm{C},\, \mathrm{C}\alpha,\, \mathrm{O}\}$. $a_n$ and $\hat{a}_n$ are the true and predicted coordinates of backbone atom $a$ at residue $n$.

The second loss penalizes deviations in local pairwise atomic distances:
% {\fontsize{9}{10.8}\selectfont
\begin{align}
\mathcal{L}_\text{2D} = \frac{
\sum_{n=1}^N \sum_{m=1}^N \sum_{a,b \in A} \mathds{1}\{d_{ab}^{nm} < 6\mathring{\mathrm{A}} \} \lVert d_{ab}^{nm} - \hat{d}_{ab}^{nm}\rVert^2
}{
\left( \sum_{n=1}^N \sum_{m=1}^N \sum_{a,b \in A} \mathds{1}\{d_{ab}^{nm} < 6\mathring{\mathrm{A}} \} \right) - N 
}
% \notag
\end{align}
% }
where $d_{ab}^{nm} = \lVert a_n - b_m \rVert$ and $\hat{d}_{ab}^{nm}$ are the true and predicted distances between atoms $a$ and $b$ in residues $n$ and $m$, respectively. The indicator function $\mathds{1}\{d_{ab}^{nm} < 6\text{\AA}\}$ limits the loss to atom pairs within 6 $\mathring{\mathrm{A}}$.

The complete training loss is given by
\begin{align}
\mathcal{L} = 
\mathcal{L}^\mathbf{R}(\theta) +
\mathcal{L}^\mathbf{X}(\theta) +
\mathds{1} \left\{ t < T/4 \right\}
\left(
\mathcal{L}_\text{bb} +
\mathcal{L}_\text{2D}
\right)
\end{align}
where $T=1$ in the case of FM.

\subsection{Model Architecture}

\begin{figure*}[!ht]
\begin{center}
\centerline{\includegraphics[width=\textwidth]{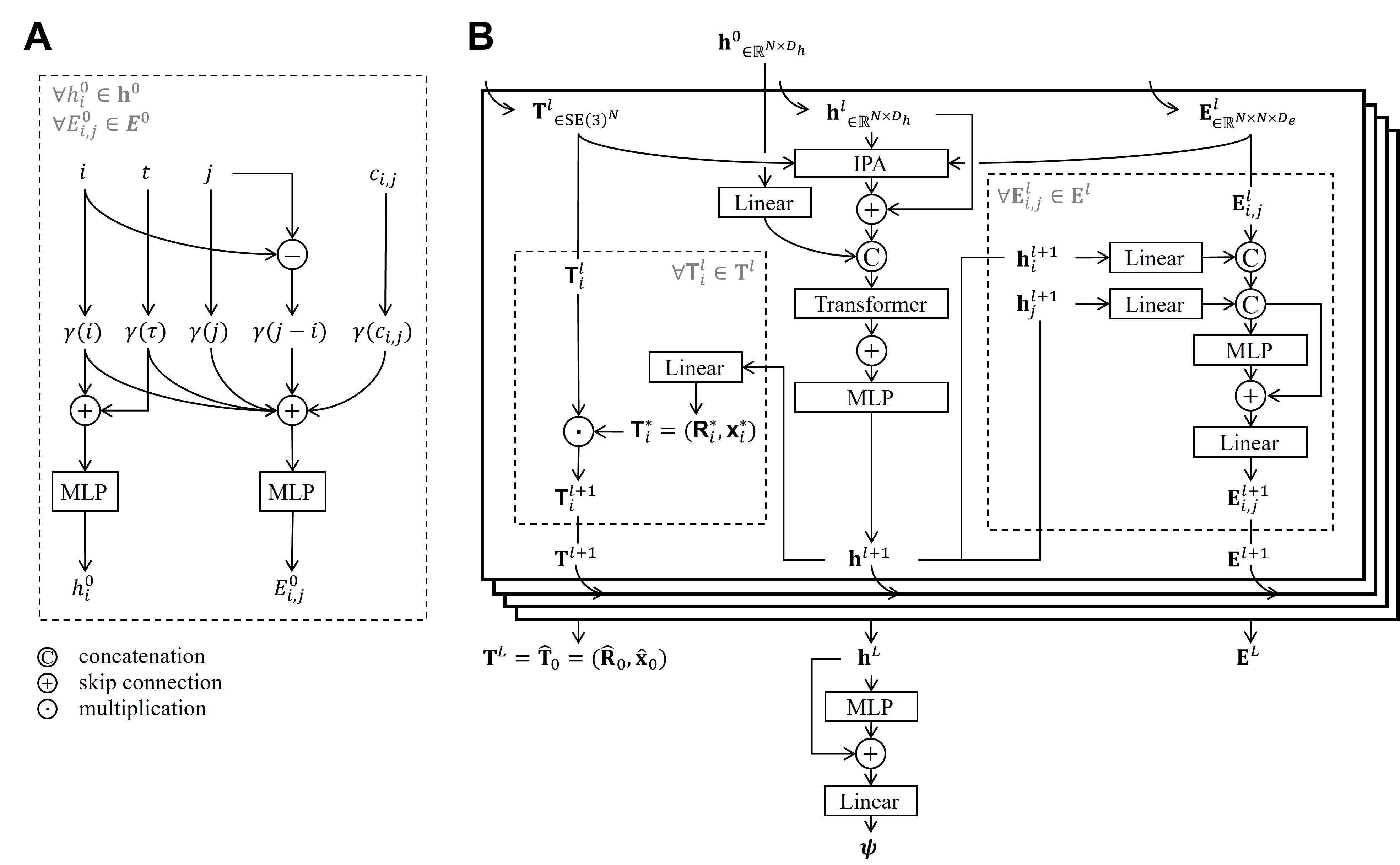}}
\vskip -0.1in
\caption{
Overview of the (\textbf{A}) embedding module and (\textbf{B}) multi-layer network architecture.
}
\label{fig:A.network}
\end{center}
\vskip -0.2in
\end{figure*}

The networks $s(\theta, t, \cdot)$ involved in SM and $v(\theta, t, \cdot)$ involved in FM models, as reviewed in \cref{appendix:B}, can share a common high-level architecture \cite{FrameDiff, FoldFlow, https://doi.org/10.48550/arxiv.2205.15019}.

\paragraph{Embeddings.}

Given node embedding dimensions $D_h$ and edge embedding dimensions $D_e$, node embeddings $\mathbf{h} \in \mathbb{R}^{N \times D_h}$ are derived from residue indices $\textbf{i} = \{1, \dots, N\}$ and time-step information $\textbf{t} = \{0, \Delta t, \dots, T\}$, while edge embeddings $\mathbf{E} \in \mathbb{R}^{N \times N \times D_e}$ integrate additional features, such as relative sequence distances $j - i$ for any $i, j \in [1,N]$ (\cref{fig:A.network}A). Self-conditioning on the predicted $\mathrm{C}_\alpha$ displacements is also applied:
\begin{align}
c_{i,j} = 
\sum_{b=1}^B 
\mathds{1} 
\left\{ 
| \bm{\mathsf{x}}_i^* - \bm{\mathsf{x}}_j^* | < v_b
\right\}
\end{align}
where $\bm{\mathsf{x}}^*$ denotes the coordinates for $\mathrm{C}_\alpha$ predicted through self-conditioning, and $v_1, \dots, v_B$ are bins spaced uniformly from 0 to $B$ angstroms. These initial features are encoded using multilayer perceptrons (MLPs) along with sinusoidal embeddings \cite{https://doi.org/10.48550/arxiv.1706.03762}.

\paragraph{Multi-layer network.}

\cref{fig:A.network}B shows the architecture of the multi-layer neural network ($L=4$ layers used in our experiments). At each layer $l$, the network takes node embeddings $\textbf{h}^l$, edge embeddings $\textbf{E}^l$, and rigid transformations $\textbf{T}^l$ as input, applying the Invariant Point Attention (IPA) introduced by \citet{AlphaFold2} to enable spatial attention. Transformer from \citet{https://doi.org/10.48550/arxiv.1706.03762} models interactions along the chain structure. The network's update procedure remains invariant under $\mathrm{SE}(3)$ transformations due to the inherent $\mathrm{SE}(3)$-invariance of the IPA.

The output $\textbf{T}^L$ from the final layer serves as the predicted frame, denoted as $\hat{\textbf{T}}_0 = (\hat{\textbf{R}}_0, \hat{\textbf{x}}_0)$. Consequently, for SM, we have the following scores predictions based on \cref{eq:SM-SCORE-R} and \cref{eq:SM-SCORE-X}:
\begin{align}
\forall s(\theta, t, \bm{\mathsf{R}}_t) \in s(\theta, t, \mathbf{R}_t), \quad
s(\theta, t, \bm{\mathsf{R}}_t)
&= \nabla_{\bm{\mathsf{R}}} \log p_{t|0} (\bm{\mathsf{R}}_t | \hat{\bm{\mathsf{R}}}_0) \\
&= \frac{\bm{\mathsf{R}}_t}{\omega(\hat{\bm{\mathsf{R}}}_0^\top \bm{\mathsf{R}}_t)}
\log \{\hat{\bm{\mathsf{R}}}_0^\top \bm{\mathsf{R}}_t\} 
\frac{\partial_\omega f(\omega(\bm{\mathsf{R}}_0^\top \bm{\mathsf{R}}_t),t)}{f(\omega(\bm{\mathsf{R}}_0^\top \bm{\mathsf{R}}_t),t)} 
\\
\forall s(\theta, t, \bm{\mathsf{x}}_t) \in s(\theta, t, \mathbf{X}_t), \quad
s(\theta, t, \bm{\mathsf{x}}_t)
&= \nabla_{\bm{\mathsf{x}}} \log p_{t|0} (\bm{\mathsf{x}}_t | \hat{\bm{\mathsf{x}}}_0) \\
&= \frac{e^{-t/2}\hat{\bm{\mathsf{x}}}_0 - \bm{\mathsf{x}}_t}{1-e^{-t}}
\end{align}
From \cref{eq:FM-R3-1} and \cref{eq:FM-SO3-2}, we have the following for FM with OT:
\begin{align}
\forall v(\theta, t, \bm{\mathsf{R}}_t) \in v(\theta, t, \mathbf{R}_t), \quad
v(\theta, t, \bm{\mathsf{R}}_t)
&= \mathbf{u}_t(\bm{\mathsf{R}}_t| \hat{\bm{\mathsf{R}}}_0, \bm{\mathsf{R}}_1 ) \\
&= \frac{\log_{\bm{\mathsf{R}}_t} ( \hat{\bm{\mathsf{R}}}_0 )}{t}
\\
\forall v(\theta, t, \bm{\mathsf{x}}_t) \in v(\theta, t, \mathbf{X}_t), \quad
v(\theta, t, \bm{\mathsf{x}}_t)
&= \mathbf{u}_t(\bm{\mathsf{x}}_t | \hat{\bm{\mathsf{x}}}_0) \\
&= \frac{ \hat{\bm{\mathsf{x}}}_0 - (1-\sigma)\bm{\mathsf{x}}_t}{1-(1-\sigma)t}
\end{align}
Torsion angle $\bm{\hat{\psi}} = \{\hat{\psi}_1, \dots, \hat{\psi}_N\} = \bm{\psi} / \lVert \bm{\psi} \rVert \in \mathrm{SO(2)}^N$ is predicted with $\textbf{h}^L$ and $\textbf{E}^L$.

\newpage
\section{Protein Families Involved in This Study}
\label{appendix:C}

\begin{figure*}[!ht]
\begin{center}
\centerline{\includegraphics[width=\textwidth]{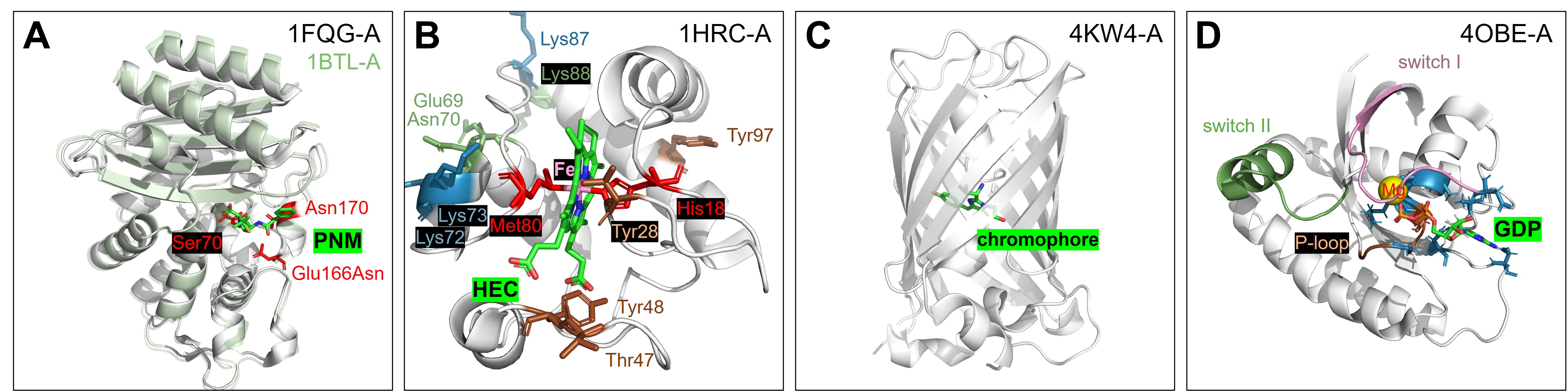}}
\vskip -0.1in
\caption{
(\textbf{A}) Overlay of WT \textit{E. coli} TEM1 (PDB: 1BTL; \citet{Jelsch1993}; green) and its E166N acylated intermediate (PDB: 1FQG; \citet{Brown2009}; white) with penicillin (PNM).
(\textbf{B}) WT \textit{E. caballus} heart cytochrome \textit{c} (PDB: 1HRC; \cite{Dickerson1967}) with heme C (HEC).
(\textbf{C}) \textit{A. victoria} GFP and its chromophore (PDB: 4KW4; \citet{Barnard2014}).
(\textbf{D}) GDP-bound \textit{H. sapiens} KRas protein (PDB: 4OBE; \cite{Hunter2014}).
}
\label{fig:A.proteins}
\end{center}
\vskip -0.2in
\end{figure*}

\subsection{$\beta$-lactamases}

$\beta$-lactamases are enzymes that deactivate $\beta$-lactam antibiotics by hydrolyzing their $\beta$-lactam ring, contributing significantly to bacterial resistance \cite{Lee2016}. 
Inhibiting these enzymes can restore antibiotic efficacy \cite{Behzadi2020}. Rapid diversification driven by the evolution of bacterial resistance makes $\beta$-lactamases ideal targets for protein modeling studies. For this study, we gathered structural data on 1,578 unique monomeric $\beta$-lactamases and their variants across Ambler classes (A, B, C, and D) from the BLDB \cite{Naas2017} and the Protein Data Bank (PDB) \cite{Berman2000}. 

Class A $\beta$-lactamases are the most prevalent, with conserved active-site residues Ser$^{70}$, Glu$^{166}$, and Asn$^{170}$ coordinating the hydrolytic water for deacylation \cite{Tooke2019, Brown2009}. \cref{fig:A.proteins}A shows the WT $\beta$-lactamase TEM1 (1BTL) alongside its acylated E166N intermediate (1FQG).

\subsection{Cytochrome \textit{c}} 

Cytochrome \textit{c} is a water-soluble protein ($\sim12$ kDa) essential for ATP synthesis in mitochondria and intrinsic apoptosis \cite{Kashyap2021, Ow2008}. It also serves as an independent marker for apoptosis in several cancers \cite{Li2001, Way2004}. Despite variations across species, its core structure and function are conserved. We obtained structural data for 498 unique cytochrome \textit{c} proteins and variants from the PDB.

\cref{fig:A.proteins}B shows horse cytochrome \textit{c}, where a hydrophobic shell surrounds the heme group, with only $\sim 7.5\%$ of the surface available for electron transfer \cite{Bushnell1990}. The hydrophobic environment and iron coordination by His$^{18}$ and Met$^{80}$ maintain a high redox potential (~260 mV) \cite{Salemme1977}. Phosphorylation occurs at Thr$^{28}$, Thr$^{47}$, Tyr$^{48}$, and Tyr$^{97}$ \cite{Httemann2011}, while Lys$^{72}$, Lys$^{73}$, and Lys$^{87}$ bind phospholipids \cite{Kagan2009}. The ATP-binding pocket involves Glu$^{69}$, Asn$^{70}$, Lys$^{88}$, and Lys$^{72}$, Lys$^{86}$, Lys$^{87}$ \cite{McIntosh1996}.

\subsection{Green fluorescent proteins (GFP)} 

GFP, first isolated from \textit{Aequorea victoria}, fluoresces green when stimulated by specific wavelengths. Its core structure is an 11-strand $\beta$-barrel enclosing the chromophore (\cref{fig:A.proteins}C) \cite{Remington2011}. Various mutants have been engineered to enhance or modify its properties, including enhanced GFP \cite{Cormack1996}, superfolder GFP \cite{Pdelacq2005}, and color variants like YFP \cite{Orm1996} and BFP \cite{Glaser2016}. We collected structural data for 448 GFPs and variants from the PDB to explore this diversity.

\subsection{Ras Proteins}

Ras proteins, a subgroup of the small GTPase superfamily, act as molecular switches, cycling between GTP-bound (active) and GDP-bound (inactive) states to regulate cell proliferation, differentiation, migration, and apoptosis \cite{Ladygina2011, Simanshu2017, Weinmann2007}. They play a key role in signaling from the cell surface to downstream pathways. Mutations that keep Ras proteins in an active state drive excessive cell growth and malignancy, making Ras inhibition a promising cancer treatment strategy \cite{Singh2002225}. We focused on the most common cancer-related Ras proteins (HRas, KRas, and NRas) \cite{Cox2002} and obtained 511 experimentally derived structures from the PDB. 

In human KRas (\cref{fig:A.proteins}D), the switch I and II regions form the key interface for effector and regulator binding \cite{Pantsar2020}. These regions are highly flexible, with conformations depending on GTP or GDP binding. Cancer-related mutations frequently occur in the P-loop and switch II \cite{Pantsar2020}.

\subsection{Sequence Length Distributions in Experimentally Derived and Generated Protein Structures}

\begin{figure*}[!ht]
\begin{center}
\centerline{\includegraphics[width=\textwidth]{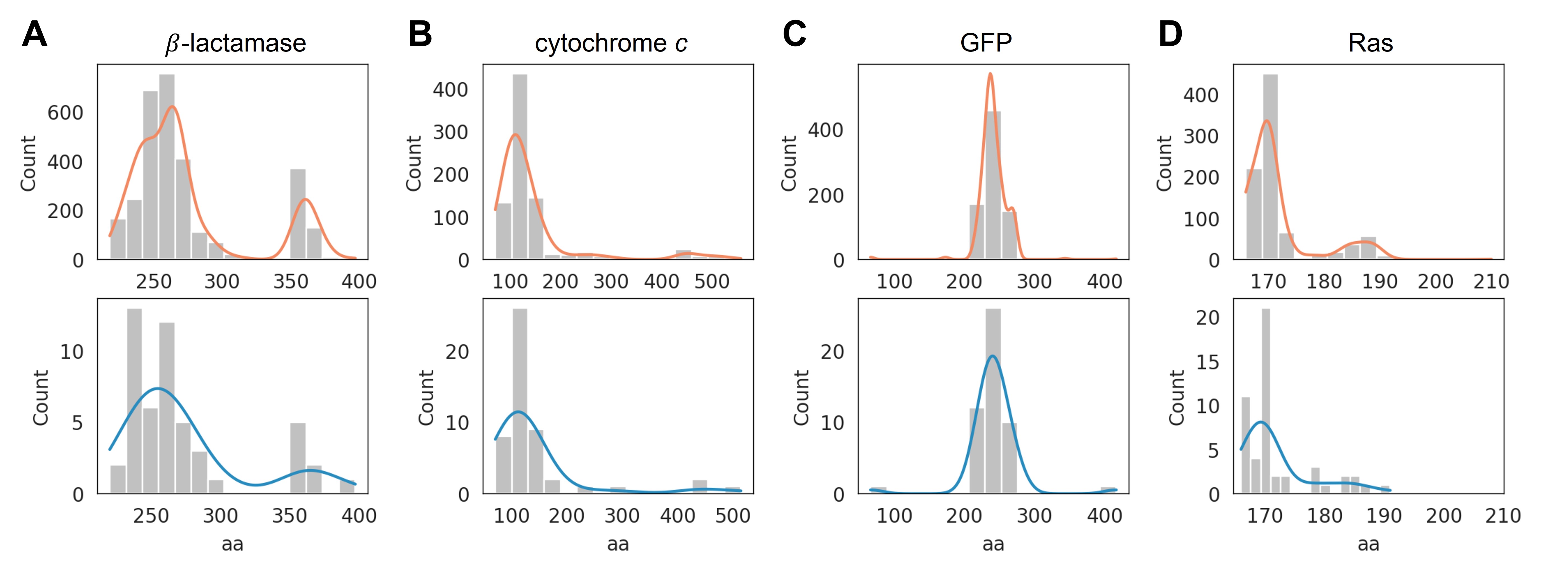}}
\vskip -0.1in
\caption{Distributions of amino acid sequence lengths (aa) for the experimentally derived protein structures used for training (top row; orange) and the 50 backbone structures generated by each model (bottom row; blue). Target sequence lengths for the generated structures were sampled from the training data distribution, shown here for (\textbf{A}) $\beta$-lactamase, (\textbf{B}) cytochrome \textit{c}, (\textbf{C}) GFP, and (\textbf{D}) Ras.
}
\label{fig:A.aadist}
\end{center}
\vskip -0.2in
\end{figure*}

\newpage
\section{Dihedral Angles $\psi$ and $\phi$ Distributions}
\label{appendix:D}

\begin{figure*}[!ht]
\begin{center}
\centerline{\includegraphics[width=\textwidth]{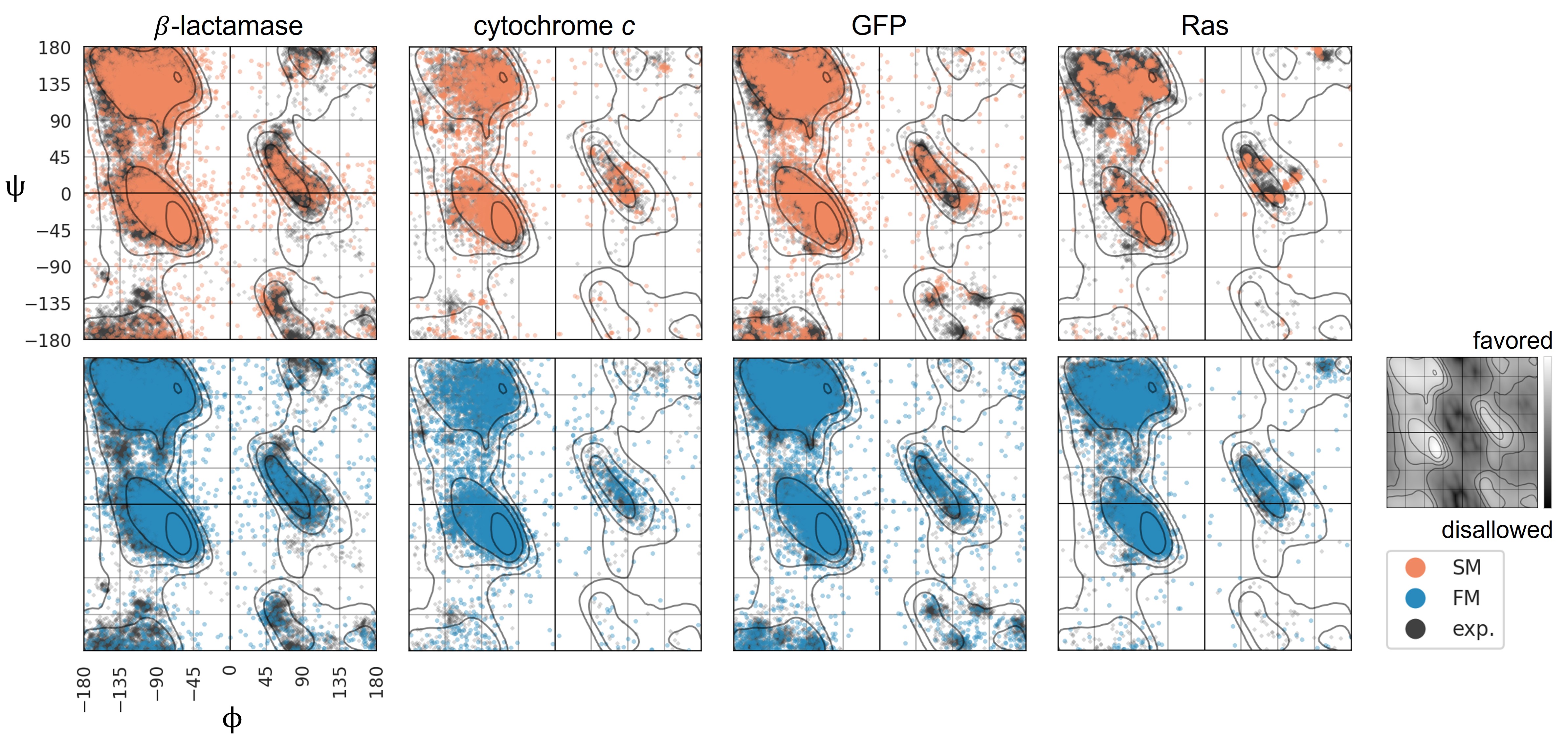}}
\vskip -0.1in
\caption{
Ramachandran plots comparing dihedral angles $\psi$ and $\phi$ distributions for generated versus experimentally derived proteins; inset shows favored (light) and disallowed (dark) regions.
}
\label{fig:A.angles}
\end{center}
\vskip -0.2in
\end{figure*}

\newpage
\section{Conserved Residue Consistency}
\label{appendix:E}

\subsection{Evolutionary Rates Mapped onto Structures}

\begin{figure*}[!ht]
\begin{center}
\centerline{\includegraphics[width=\textwidth]{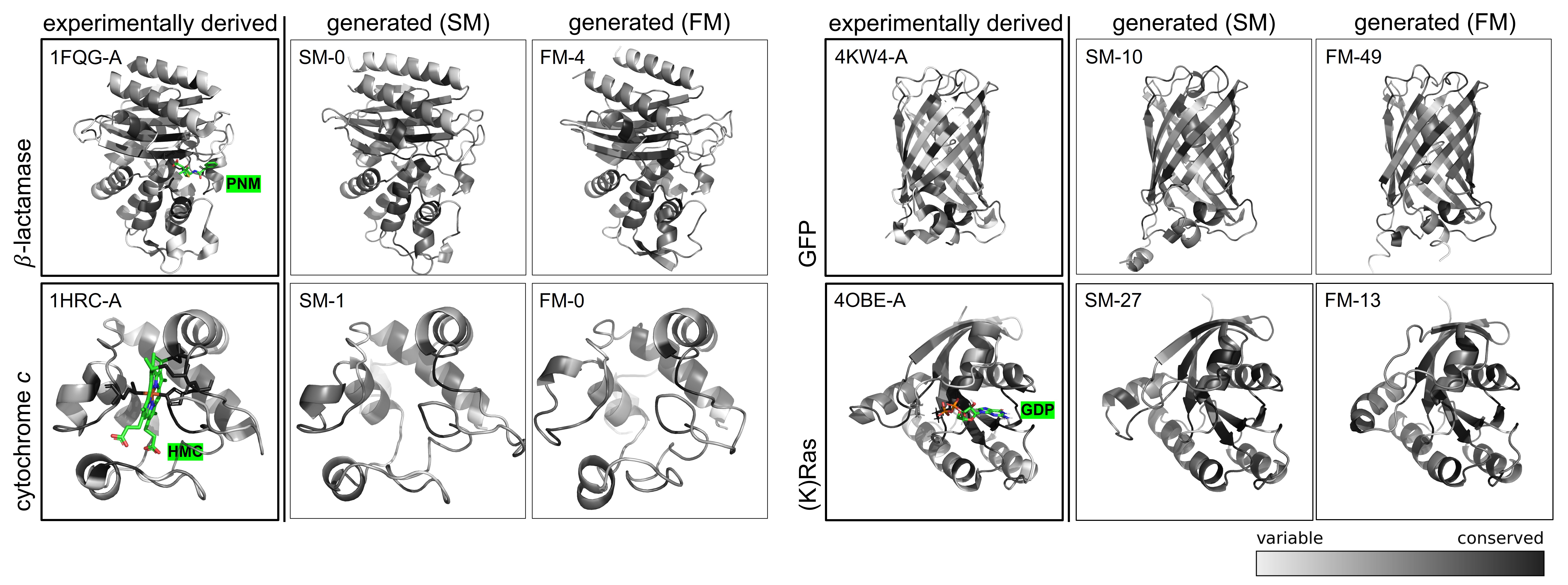}}
\vskip -0.1in
\caption{
Normalized evolutionary rates mapped onto structures, with white for rapidly evolving positions and black for conserved ones. Experimentally derived structures are highlighted with bold outlines. Refer to \cref{fig:A.proteins} and \cref{fig:A.mddocking} for the relevant key residues and the locations of the ligand-binding pockets (represented by PNM, HMC and GDP).
}
\label{fig:A.resevs}
\end{center}
\vskip -0.2in
\end{figure*}

\newpage
\subsection{Evolutionary Rates Mapped onto Sequences}

\begin{figure*}[!ht]
\begin{center}
\centerline{\includegraphics[width=\textwidth]{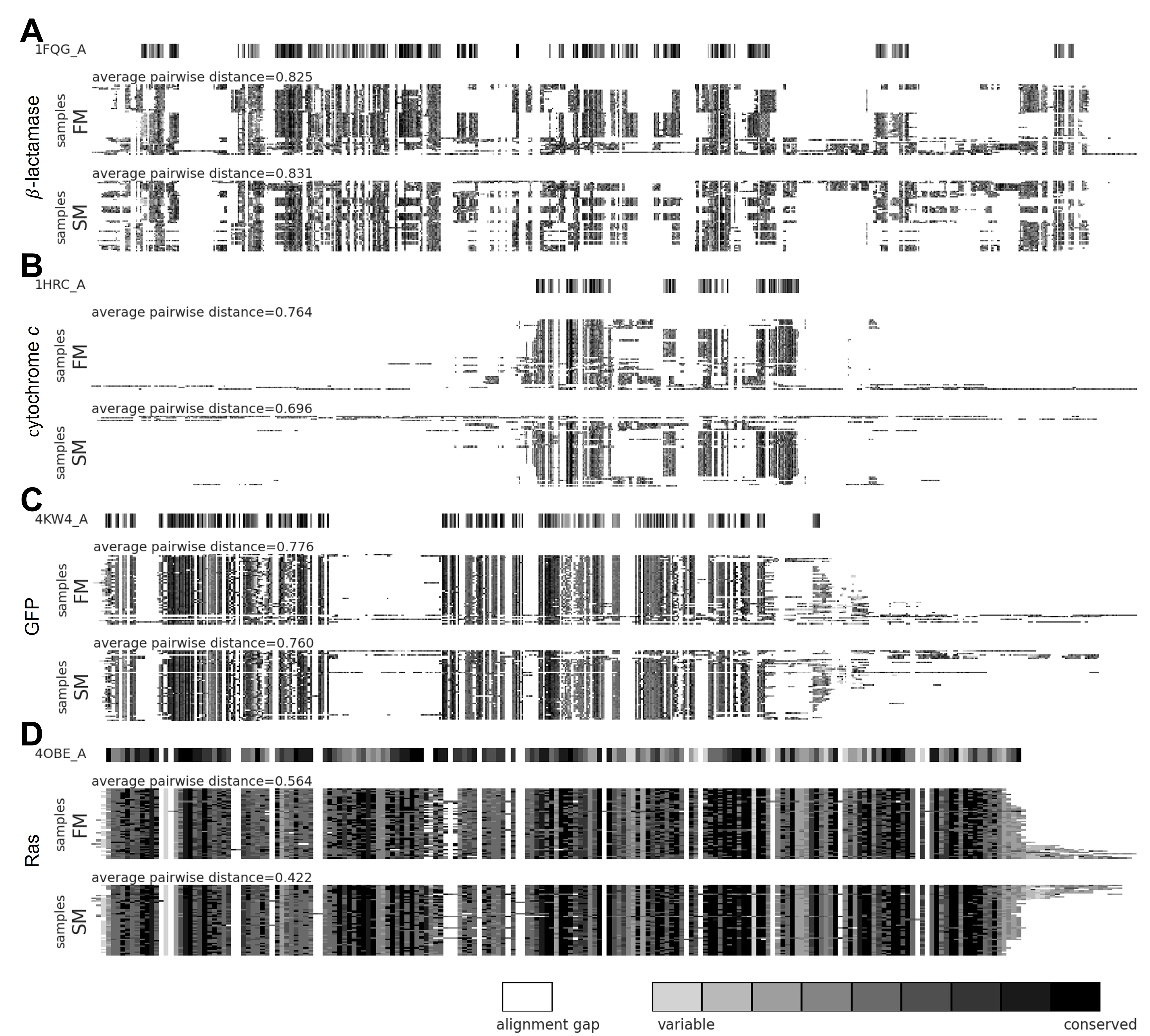}}
\vskip -0.1in
\caption{
Multiple sequence alignment of generated samples and experimentally derived reference proteins (1FQG, 1HRC, 4XWA, 4O8E) using Clustal Omega, with normalized evolutionary rates overlaid. Rapidly evolving positions are highlighted in light gray, conserved regions in black, and alignment gaps in white. 
In addition, a comparative analysis of variability is presented: 
average pairwise distances, derived from the Clustal Omega distance matrix, denote greater variability when larger values are observed.
}
\label{fig:A.resevo}
\end{center}
\vskip -0.2in
\end{figure*}

\begin{figure*}[!ht]
\begin{center}
\vskip 0.2in
\centerline{\includegraphics[width=.7\textwidth]{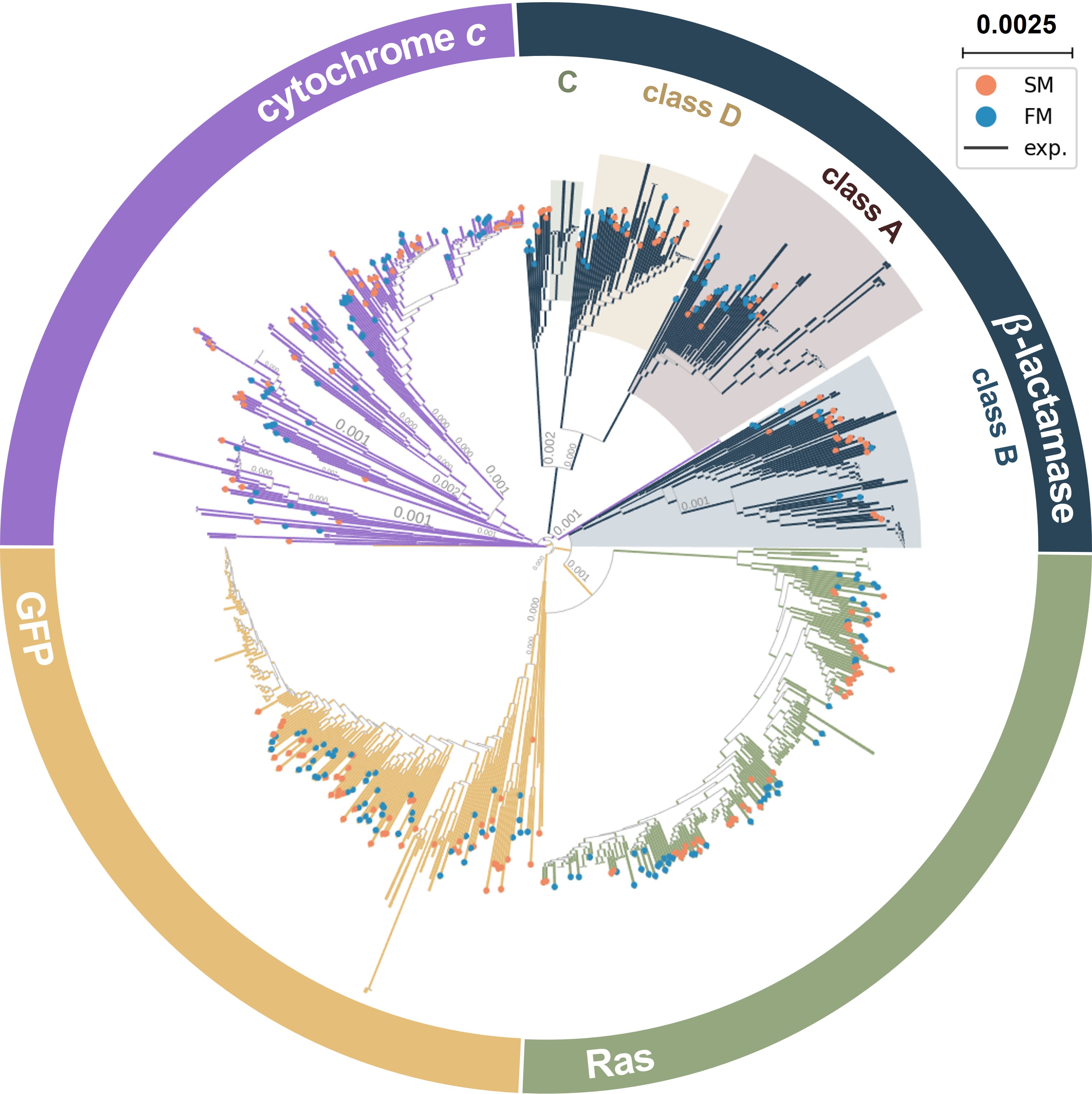}}
\vskip 0.2in % \vskip -0.1in
\caption{
Summary structural phylogenetic tree constructed using the $Q_{\text{score}}$ and the 3Di alphabet. Branch colors distinguish families, with orange and blue nodes representing SM- and FM-generated structures, respectively. In $\beta$-lactamases, distinct Ambler classes are differentiated by unique background colors.
}
\label{fig:A.phylo.summary}
\end{center}
\vskip -0.2in
\end{figure*}

\cref{fig:A.resevo} shows that, except for $\beta$‑lactamase, the FM-generated sequences show a slightly higher average pairwise distance than those of the SM, indicating greater diversity. One possible explanation is that the experimentally derived $\beta$‑lactamase structures used for training (from four distinct Ambler classes) have more variability than the other three protein families (as also reflected in \cref{fig:A.phylo.summary}) and that SM is more sensitive to structural variability.

\newpage
\section{Summary Structural Phylogenetic Tree}\label{appendix:F}

\newpage
\section{Structure-informed Trees versus Sequence-based Phylogenetic Trees}
\label{appendix:G}

\paragraph{Retrieving known taxonomic lineages.}

Following \citet{Moi2023}, we retrieved taxonomic lineages for each experimentally derived sequence and structure within every protein family using the UniProt API \cite{Patient2008}, assuming that most genes evolve in a manner that mirrors the species tree with only occasional instances of gene loss or duplication.

\begin{figure*}[!ht]
\begin{center}
\centerline{\includegraphics[width=\textwidth]{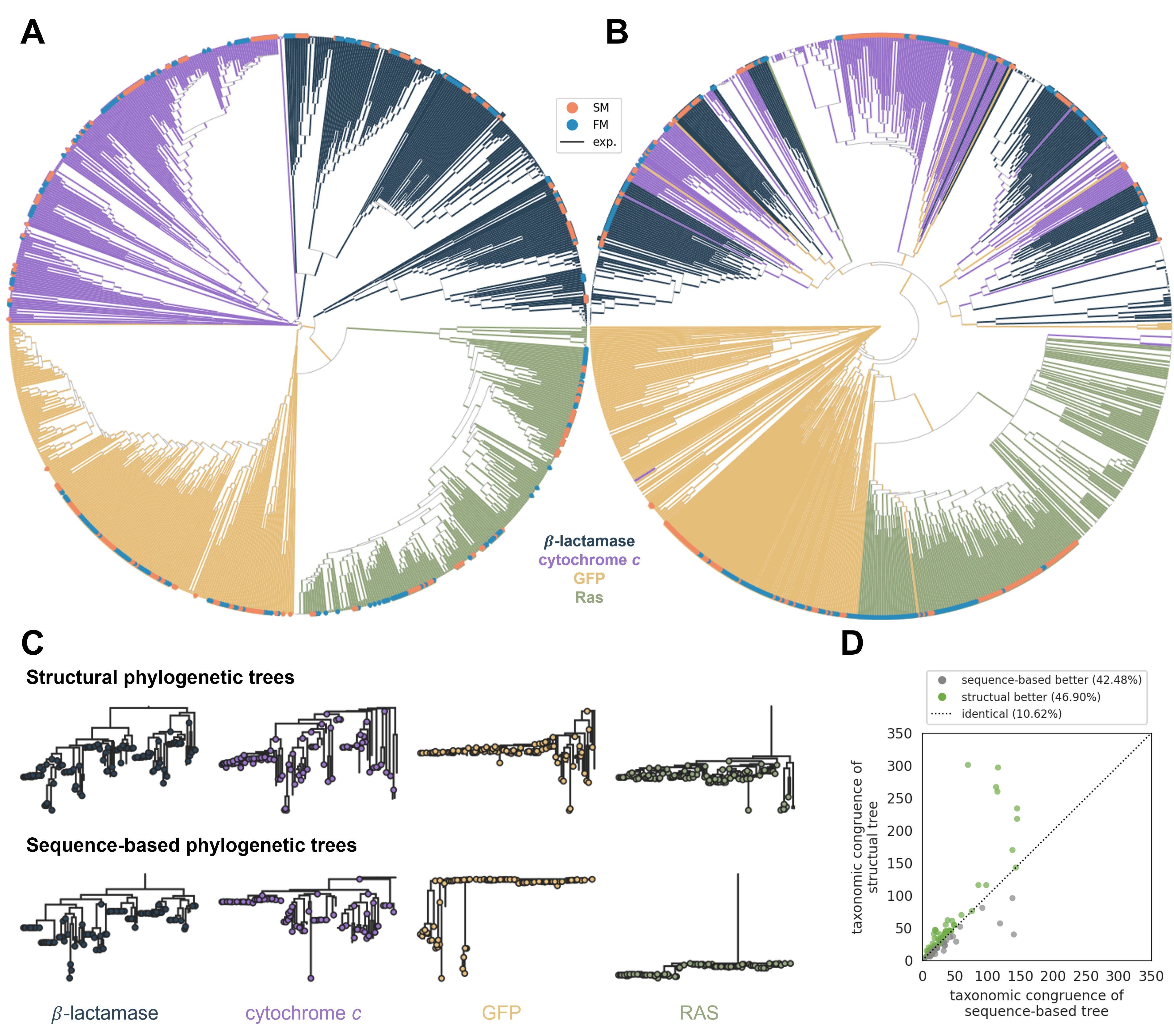}}
% \vskip -0.1in
\caption{
(\textbf{A}) Ultrametric summary structural phylogenetic tree constructed using the $Q_{\text{score}}$ and the 3Di alphabet. 
(\textbf{B}) Ultrametric sequence-based phylogenetic tree constructed using the Clustal Omega and the FastTree pipeline. 
Different protein families are differentiated by distinct colored branches. Nodes in orange and blue represent structures generated by SM and FM, respectively.
(\textbf{C}) Phylogenetic trees of distinct protein families. Top: summary structural phylogenetic tree constructed using the $Q_{\text{score}}$ metric and the 3Di alphabet. Bottom: sequence-based phylogenetic tree inferred using the Clustal Omega and FastTree pipeline.
(\textbf{D}) Taxonomic congruence score for each node in the sequence and structure trees. On average, structural trees exhibit higher taxonomic congruence than sequence-based trees.
}
\label{fig:A.phylo}
\end{center}
\vskip -0.2in
\end{figure*}

\paragraph{Using $Q_\text{score}$ in structural phylogenetics.}

For any two structures with $N_1$ and $N_2$ residues, $Q_\text{score}$ is computed with TM-align \cite{Zhang2005} as:
\begin{align}
Q_\text{score}=
% \overbrace{
\frac{N_\text{align}^2}{N_1N_2}
% }^{Q_\text{length}} 
\times 
% \overbrace{
\frac{1}{1+\left(\frac{\text{RMSD}}{R_0}\right)^2}
% }^{Q_\text{shape}} 
\end{align}
where $N_\text{align}$ is the number of aligned residues, RMSD is the root-mean-square deviation of atomic positions, and $R_0$ (set to $4 \mathring{\text{A}}$) balances the contributions of RMSD and $N_\text{align}$.

\paragraph{Taxonomic congruence score (TCS).} 

\citet{Tan2015} proposed the use of the TCS to assess how well a phylogenetic tree’s topology agrees with the known taxonomy, arguing that TCS is an unbiased measure of tree quality. Typically, trees with higher average TCS (structure trees in \cref{fig:A.phylo}D) are considered to have more accurate topologies. In this study, we evaluated the congruence between the given trees and the established taxonomy lineages derived by UniProt. Subsequently, we provide a brief overview of the bottom-up TCS implementation as described by \citet{Moi2023}:

For any node $x$ in the tree, $s(x)$ is the set of taxonomic lineage labels present in the subtree rooted at $x$. If $x$ is a leaf, then $s(x)$ is defined to be the set of lineage labels (its taxonomic classification) for that leaf. If $x$ is an internal node with children $\{y_1, \cdots, y_N\}$, then $s(x) = \bigcap_{i=1}^N s(i)$.

For each node $x$, the function $C(x)$ quantifies the congruence of that node’s grouping with taxonomy: $C(x) = |s(x)| + |s(p)|$, where $p$ is the parent nodes of $x$, and $|\cdot|$ denotes the size of a set.

The overall taxonomic congruence score for the entire tree is obtained by summing the contributions of all leaves. To compare congruence across trees of different sizes, the raw total score is typically normalized.

\newpage
\section{Evaluation of Side-Chain Homology Modeling}
\label{appendix:H}

We used homology modeling to add side-chains to the generated protein backbones, evaluating them using PROCHECK \cite{Laskowski1993, Laskowski1996} and WHAT\_CHECK \cite{Hooft1996} to correct or exclude those not meeting the expectations. Specifically:

\paragraph{Planarity.}

Planar side-chains, such as those in phenylalanine, tyrosine, tryptophan, and histidine, are essential for stability and function. Conformations lacking expected planarity were discarded.

\paragraph{Asparagine, glutamine, histidine flips.}

Asparagine, glutamine, and histidine side-chains can experience terminal flips, altering key interactions. WHAT\_CHECK was used to evaluate and, if necessary, adjust side-chain orientations to have more stable interactions.

\paragraph{Torsion angles.}

Side-chain torsion angles ($\chi$ angles) were assessed, focusing on $\chi_1$ (rotation around $\mathrm{C}_\alpha$ to the first side-chain atom) and $\chi_2$ (rotation to the second side-chain atom) to prevent spatial clashes. Conformations in uncommon $\chi$-angle regions were excluded.

\paragraph{Bond lengths and angles.}

Unusual bond lengths and angles may indicate strain and modeling errors, potentially disrupting interactions. Conformations with such issues were discarded.

\paragraph{Other parameters.}

Side-chains with abnormal torsion angles, atypical aromatic bonding angles, or unusual proline puckering were also discarded.

\newpage
\section{Molecular Dynamics and Blind Docking Simulations}
\label{appendix:I}

\begin{figure*}[!ht]
\begin{center}
\centerline{\includegraphics[width=\textwidth]{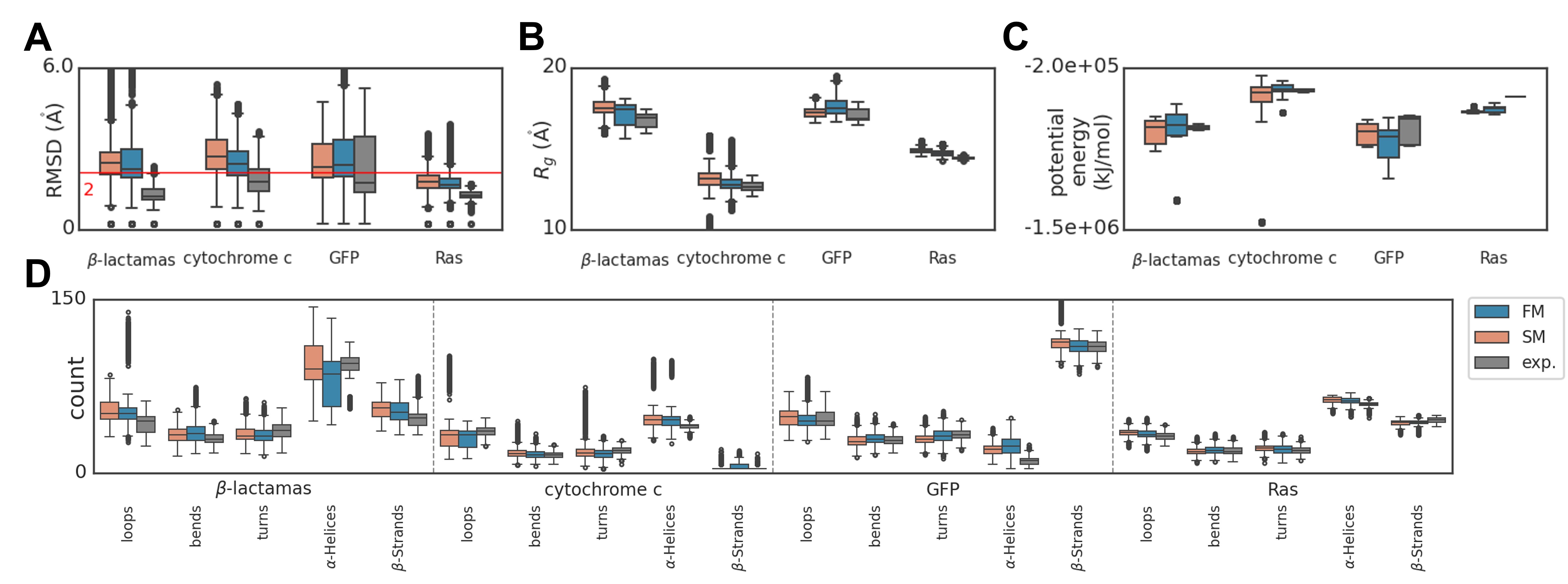}}
\vskip -0.1in
\caption{
Stability assessment of MD simulations across proteins using various metrics. Distributions of (\textbf{A}) RMSD, (\textbf{B}) radius of gyration ($R_g$), (\textbf{C}) potential energy, and (\textbf{D}) secondary structure counts throughout the simulation. Interquartile ranges and whiskers show metric variation; high-quality structures have medians close to experimentally derived values with narrow ranges.
}
\label{fig:A.mdmetrics}
\end{center}
\vskip -0.2in
\end{figure*}

For experimentally derived structures, crystallographic water and unnecessary small molecules were removed. For generated structures, missing side-chains were added via homology modeling (\cref{sec:md_homo_side}). After adding hydrogen atoms using Reduce2 \cite{GrosseKunstleve2002} and confirming no missing atoms, each protein was centered at the origin. 

Simulations were performed with GROMACS \cite{Abraham2015}, using the all-atom CHARMM36 force field (July 2022 version) \cite{Vanommeslaeghe2009, Vanommeslaeghe2012, Vanommeslaeghe2012-2, Yu2012, SoterasGutirrez2016}.

\subsection{Molecular Dynamics Setup for Stability Assessment of Generated Structures}
\label{appendix:I1}

Proteins were placed in an octahedral simulation box with a minimum distance of 1.5 nm between the protein and the box boundaries. Prior to solvation, energy minimization was performed in vacuum using the steepest descent method (max 30,000 steps, step size 0.01 nm, convergence 2 kJ/(mol$\cdot$nm)) to resolve steric clashes and geometric inconsistencies. Neighbor searching used a grid-based method with a search radius of 1.2 nm.

In accordance with GROMACS 2024 documentation, we applied the following configurations in the MD parameter (.mdp) files. Van der Waals interactions were handled using a cutoff method, while long-range electrostatic interactions were calculated using the Particle Mesh Ewald (PME) method.

{\footnotesize
\begin{verbatim}
constraints     = h-bonds
cutoff-scheme   = Verlet
vdwtype         = cutoff
vdw-modifier    = force-switch
rlist           = 1.2
rvdw            = 1.2
rvdw-switch     = 1.0
coulombtype     = PME
rcoulomb        = 1.2
DispCorr        = no
\end{verbatim}
}

After solvating the system with water using the TIP3P model, we added Na$^+$ and Cl$^-$ ions to achieve a physiological concentration of 150 mM and to neutralize the system's total charge. Energy minimization was then conducted to resolve steric clashes and optimize the geometry, with potential energy and maximum force monitored to ensure they reached acceptable thresholds.

The next step involved equilibrating the solvent and ions around the protein. We chose the leap-frog integrator for the simulations and applied the LINCS algorithm to constrain hydrogen bonds. Equilibration involved two stages. In the first stage, we performed a 500 ps NVT\footnote{Constant number of particles, volume, and temperature.} equilibration (250,000 steps with a 2 fs time step). Temperature control was managed using the V-rescale thermostat, with the system divided into two groups: (1) protein and (2) water + ions, both set to a target temperature of 310 K to simulate physiological conditions. In the second stage, we carried out a 500 ps NPT \footnote{Constant number of particles, pressure, and temperature.} equilibration with pressure coupling enabled. The pressure was regulated using the C-rescale method with isotropic coupling. The target pressure was 1.0 bar, with a compressibility of $4.5 \times 10^{-5}$ bar$^{-1}$ and a pressure coupling time constant of 0.5 ps.

Following equilibration, we conducted a 10 ns production simulation (5,000,000 steps with a 2 fs time step), during which all position restraints were removed. This allowed us to observe and analyze the system's dynamic behavior over time, in order to access its stability. Full details of the MD parameter files can be found in Software and Data.

\subsection{Molecular Dynamics Setup for Conformational Analysis of Protein-Ligand Complexes}
\label{appendix:I2}

The receptor and ligand were saved as separate coordinate files to prepare their respective topologies. The receptor topology was prepared as in \cref{appendix:I1}. For the ligand, hydrogen atoms were added using OpenBabel \cite{OBoyle2011}, and topology was generated via the CGenFF server \cite{Vanommeslaeghe2009, Vanommeslaeghe2012, Vanommeslaeghe2012-2}. The receptor and ligand topologies, along with force-field-compatible coordinate files, were then combined to construct the complete complex system.

The MD workflow for complexes followed the same steps in \cref{appendix:I1}. Complexes were placed in an octahedral simulation box, energy-minimized in vacuum, solvated in water, and neutralized with Na$^+$ and Cl$^-$ ions to 150 mM. A second energy minimization was then performed on the solvated system.

During equilibration, positional restraints were applied to the ligand to prevent unnecessary displacement in the initial stages of the simulation. Additionally, to minimize interference from temperature fluctuations of the ligand on the overall simulation, we defined two temperature coupling groups: (1) the receptor and ligand as one group, and (2) the solvent and ions as the other. Other equilibration settings followed \cref{appendix:I1}.

\begin{figure*}[!ht]
\begin{center}
\centerline{\includegraphics[width=\textwidth]{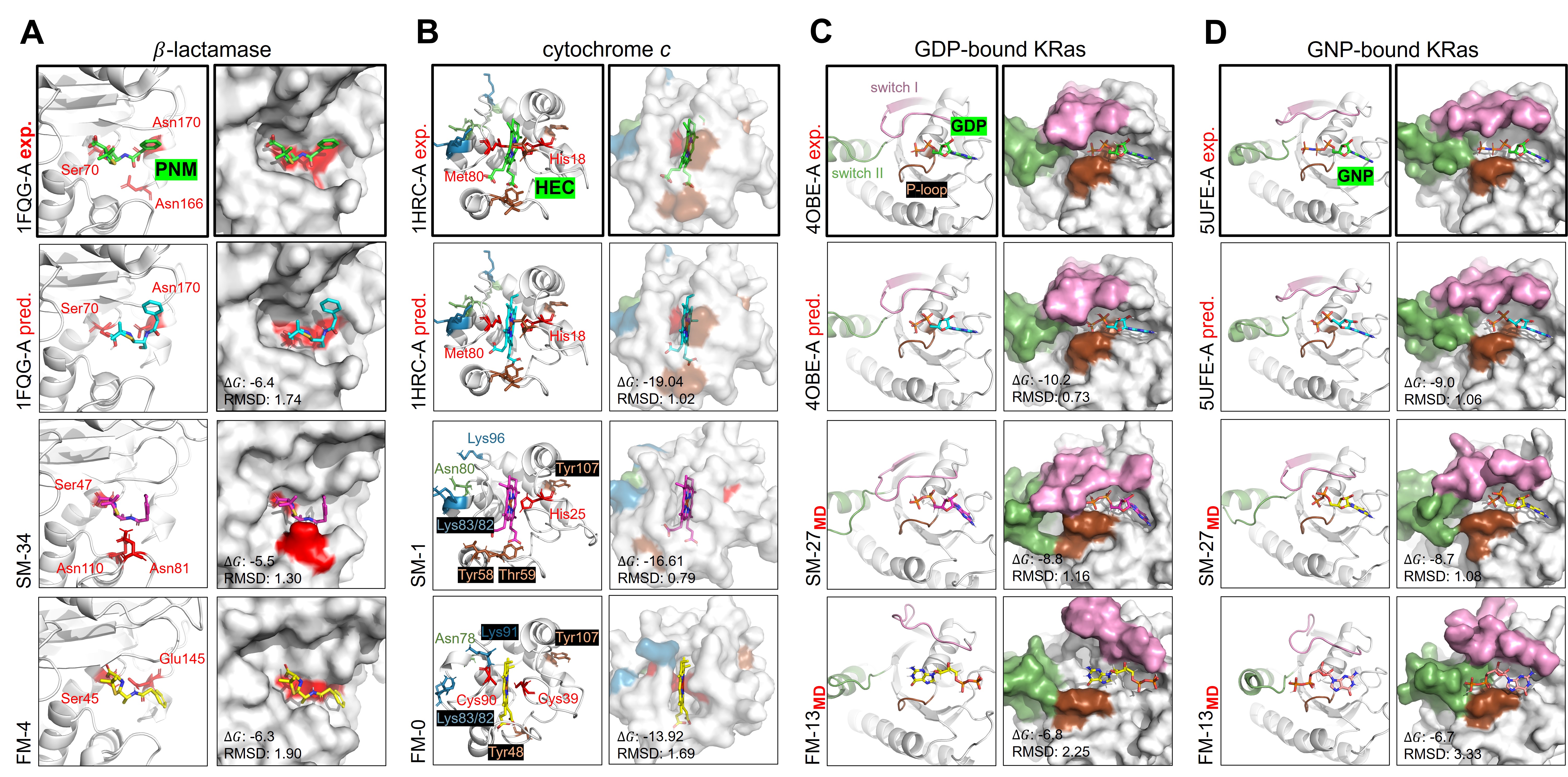}}
% \vskip -0.1in
\caption{
Comparison of experimentally derived binding modes (bold boxes; first row) with predictions from blind docking simulations for receptors binding to family-specific ligands. Ligands are colored as green (ground truth), cyan (predicted using experimentally derived receptors), magenta (predicted using SM samples), and yellow (predicted using FM samples). $\Delta G$ (kcal/mol) and RMSD ($\text{\AA}$) quantify binding affinities and deviations from the experimental poses. ``MD'' labels protein-ligand complex MD simulations after docking.
}
\label{fig:A.mddocking}
\end{center}
\vskip -0.2in
\end{figure*}

After equilibration, restraints were removed, and a 10 ns production simulation was conducted to analyze the dynamic behavior and conformational changes in the complexes.

\subsection{Protein-ligand Blind Docking}
\label{appendix:I3}

Similarly, crystallographic water and unwanted molecules were removed from experimentally derived structures, and missing side-chains were added to generated structures via homology modeling. Receptor structures were prepared using AutoDock Tools \cite{Morris2009}, with polar hydrogens added, Kollman charges assigned, and any missing atoms repaired. For receptors within the same family, we prepared a shared ligand file, adding hydrogen atoms and assigning Gasteiger charges. A large grid box, typically 80 to 110 $\mathring{\text{A}}$ per side, was defined to cover the entire protein surface.

Using these settings, we performed blind docking with AutoDock Vina \cite{Trott2009, Eberhardt2021}, generating up to 25 binding modes with a maximum energy difference of 5 kcal/mol and an exhaustiveness level of 20. The binding mode with the lowest binding free energy was selected as the final result.

\newpage
\section{Applicability and Limitations of Deep Generative Protein Design Guidelines}
\label{appendix:J}

\begin{itemize}
\item It is hard to apply strict physical constraints to very flexible proteins in generative process. In addition, using only steric exclusions without considering environmental factors such as the lipid bilayer in membrane proteins or interfaces in large assemblies can make the designs less accurate \cite{Winnifrith2024}.

\item If a target protein's stability or fold depends critically on bound cofactors or oligomeric assembly, designing sequences in isolation can be unreliable. According to \citet{Krishna2024}, generating and screening candidate sequences without accounting for these interactions will often fail to reproduce the native structure or stability observed in the full cofactor- or multimer-associated complex.

\item For truly new folds or functions, there are no known conserved residues to guide design. In families where conserved sites are spread out or have moved over evolution, keeping those sites can block creative changes. In practice, the design of novel proteins remains a low-success ``attritional'' problem with success only in rare, isolated cases \cite{Greener2018}.

\item Standard force fields often do not model metal ions, sugar attachments, or membrane effects accurately. As a result, refinement steps may produce structures that differ from what actually happens in the lab.

\item For novel protein folds with no known homologs, template-based modeling or comparison is difficult. Even state-of-the-art predictors admit substantially reduced accuracy when no homologous structure exists \cite{AlphaFold2}.

\item Conventional MD simulations are limited to relatively short timescales due to computational cost. For example, simulating $\sim50,000$ atoms (a modest protein) for $\sim1$ $\mu$s can take days on a GPU \cite{Hollingsworth2018}. Moreover, simulations of systems with membranes, metal centers, or covalent modifications often suffer from force-field artifacts or setup uncertainties.

\item Docking methods that do not model backbone flexibility often cannot accommodate the large backbone and side-chain rearrangements required for binding, leading to unreliable predictions \cite{Lexa2012}.
\end{itemize}

\newpage
\section{Generated Structures}
\label{appendix:K}

\begin{figure*}[!ht]
\begin{center}
\centerline{\includegraphics[width=\textwidth]{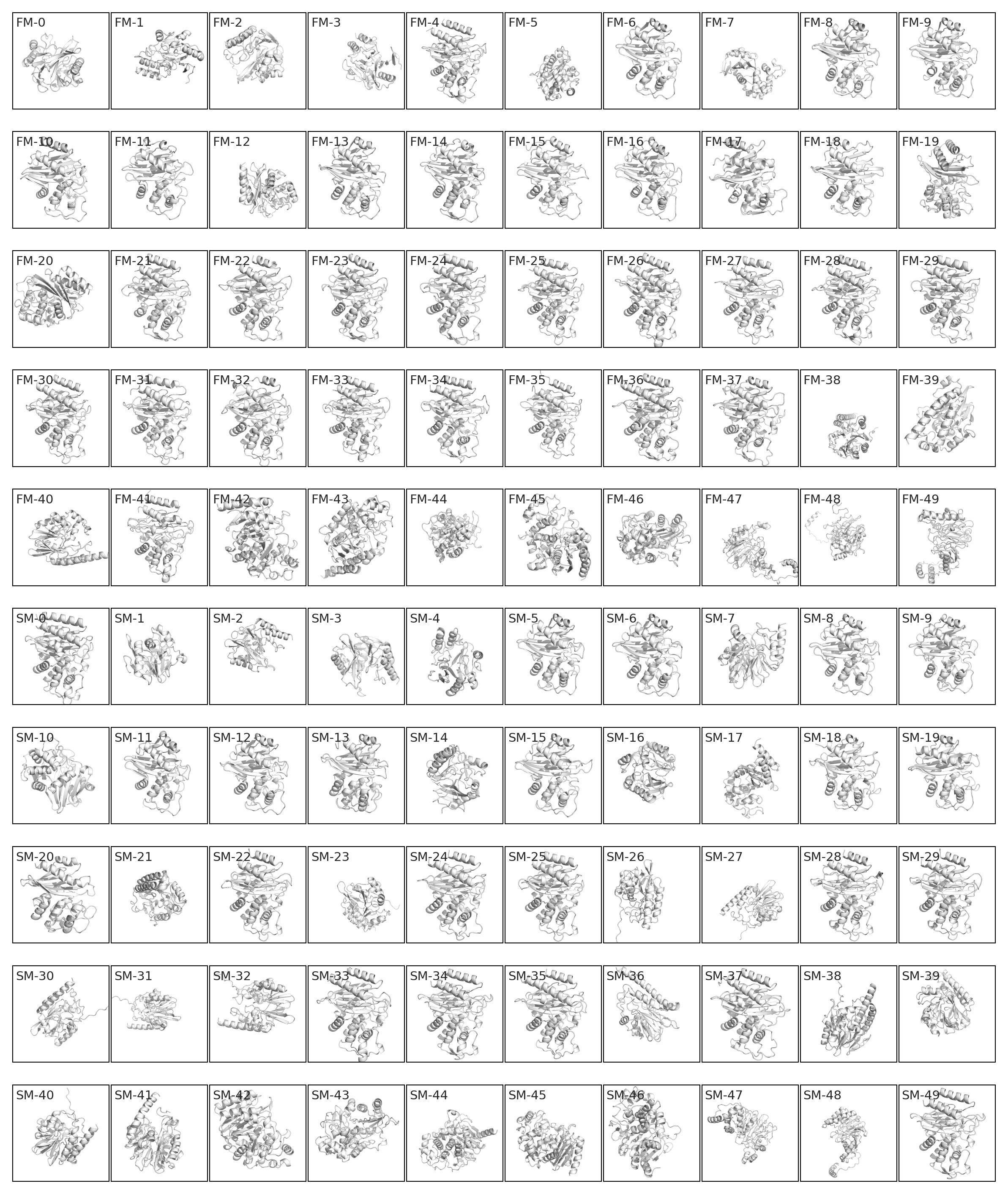}}
\caption{
50 $\beta$-lactamase-like protein backbones generated using score matching and 50 using flow matching.
}
\label{fig:X1}
\end{center}
\vskip -0.2in
\end{figure*}

\begin{figure*}[!ht]
\begin{center}
\centerline{\includegraphics[width=\textwidth]{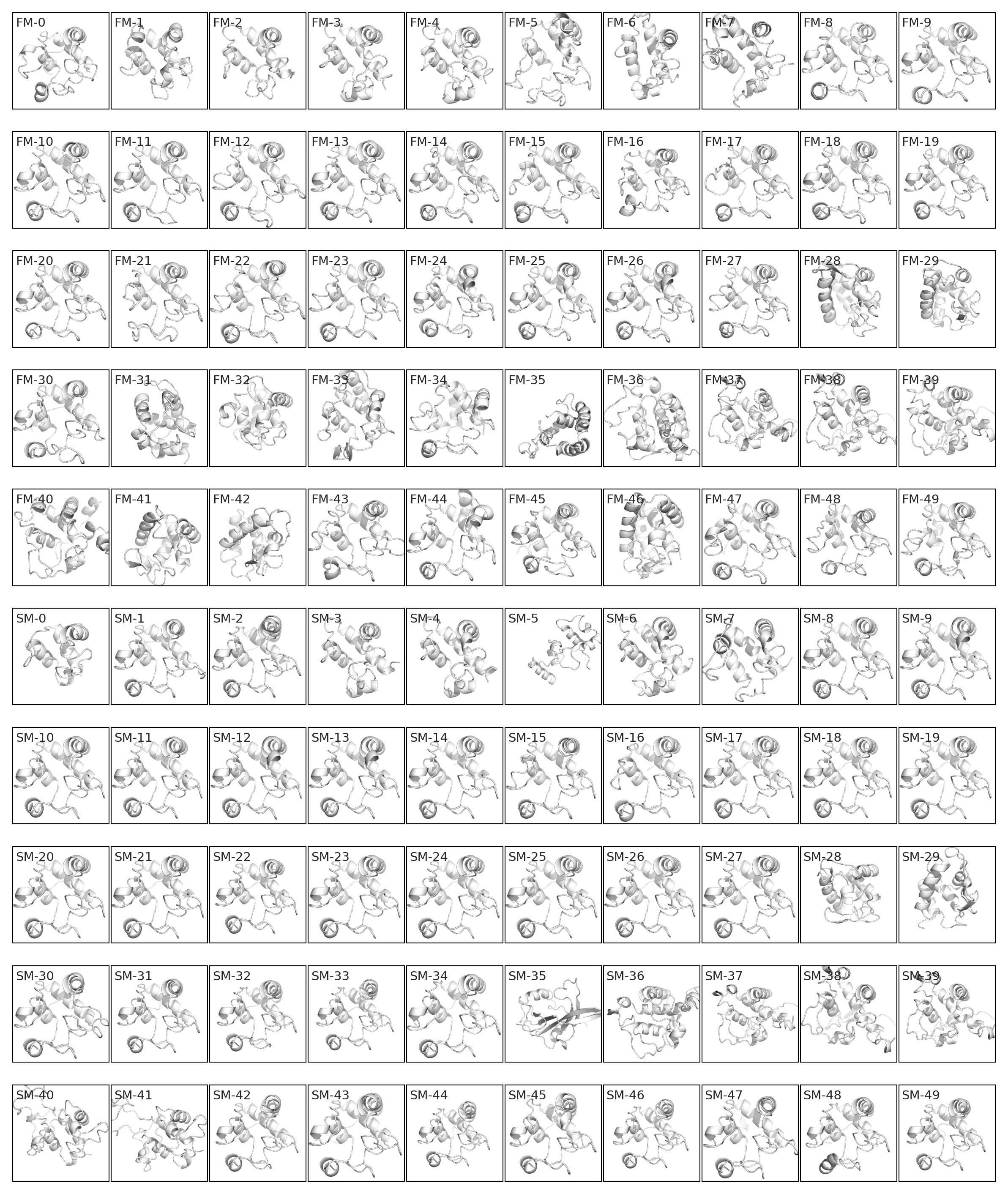}}
\caption{
50 cytochrome \textit{c}-like protein backbones generated using score matching and 50 using flow matching.
}
\label{fig:X2}
\end{center}
\vskip -0.2in
\end{figure*}

\begin{figure*}[!ht]
\begin{center}
\centerline{\includegraphics[width=\textwidth]{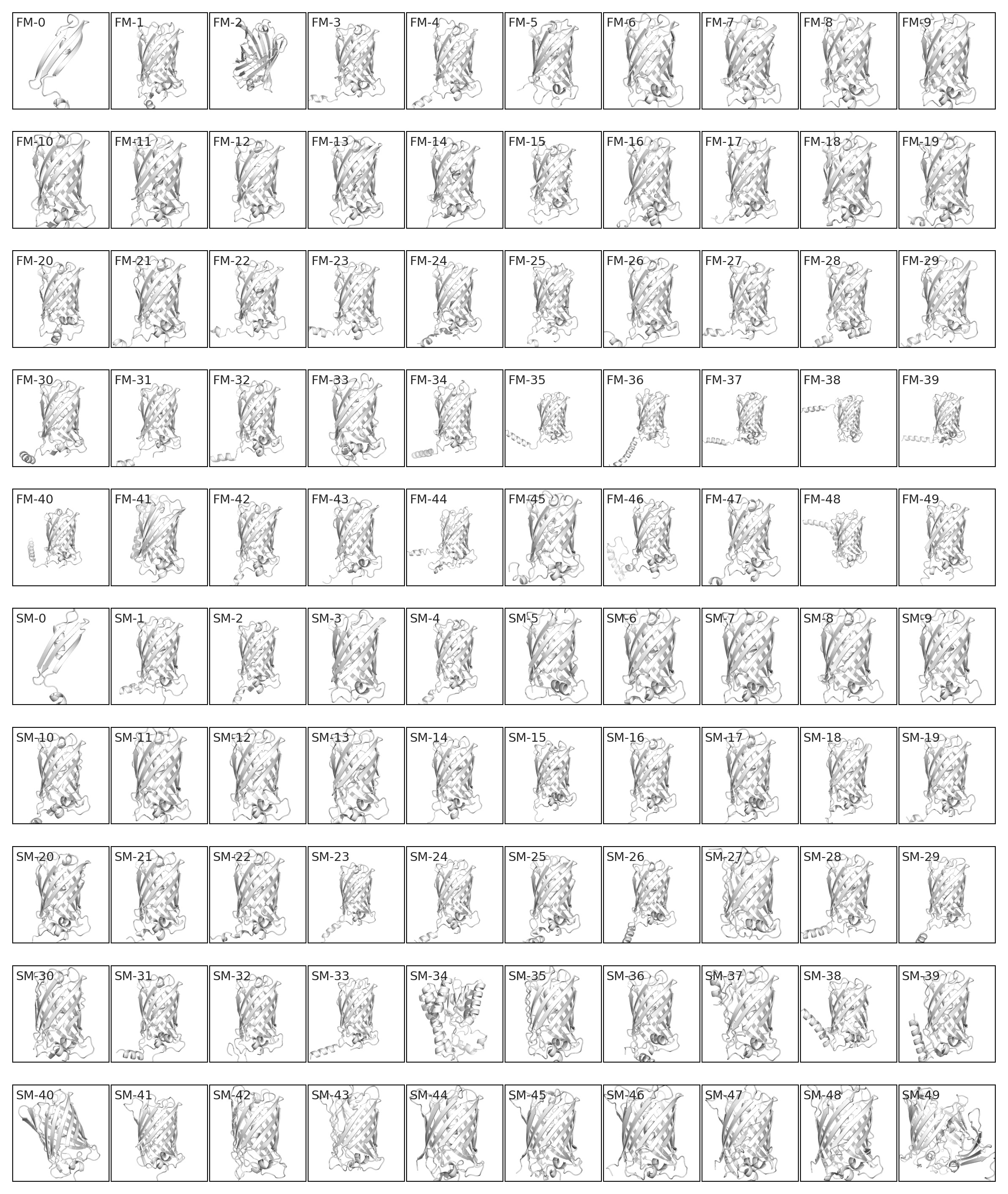}}
\caption{
50 GDP-like protein backbones generated using score matching and 50 using flow matching.
}
\label{fig:X3}
\end{center}
\vskip -0.2in
\end{figure*}

\begin{figure*}[!ht]
\begin{center}
\centerline{\includegraphics[width=\textwidth]{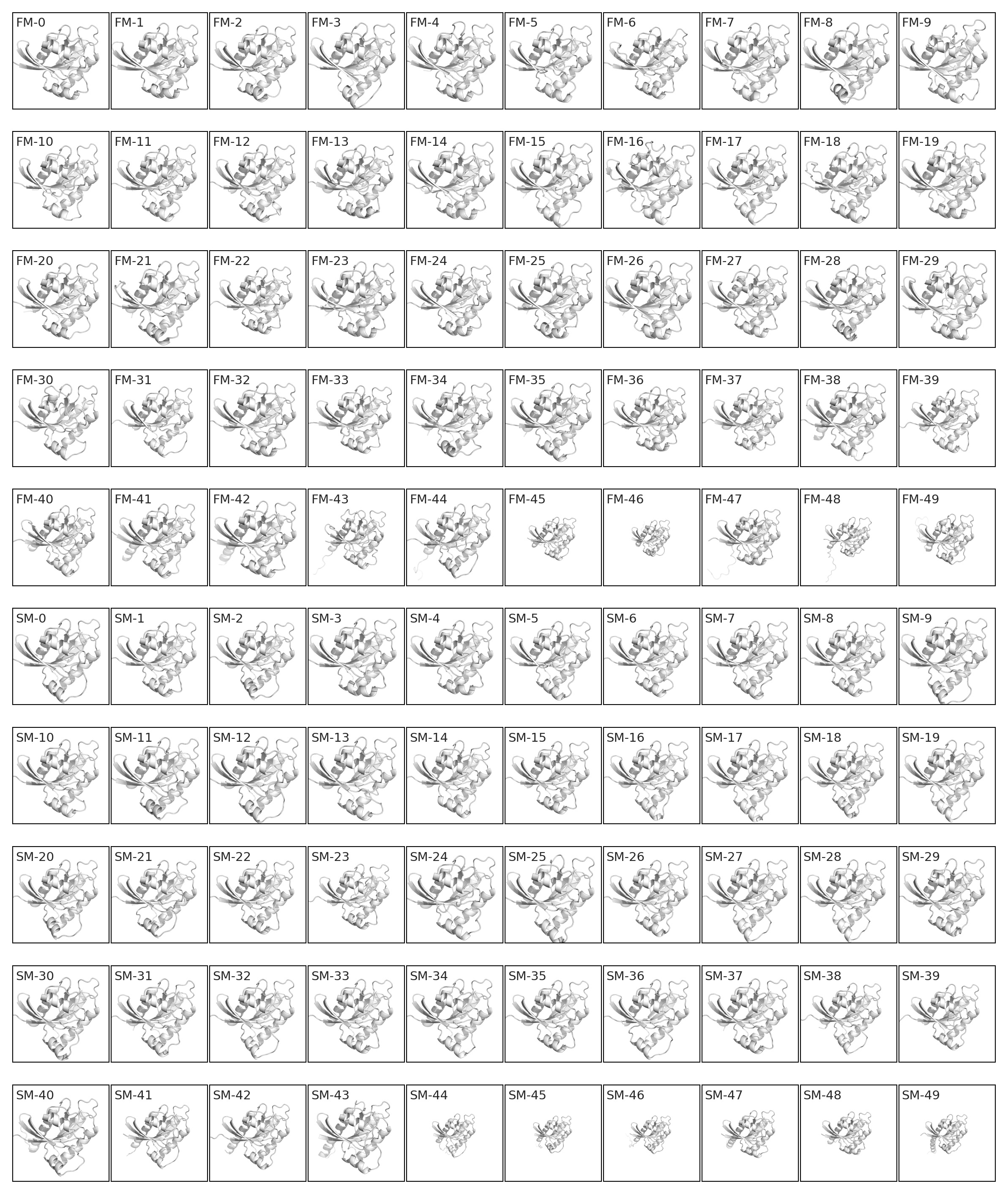}}
\caption{
50 Ras-like protein backbones generated using score matching and 50 using flow matching.
}
\label{fig:X4}
\end{center}
\vskip -0.2in
\end{figure*}

\end{document}